\documentclass[pra,aps,superscriptaddress,nofootinbib,twocolumn]{revtex4}

\usepackage{mathrsfs}
\usepackage{amsfonts}
\usepackage{amssymb}
\usepackage{amsmath}
\usepackage{graphicx}
\usepackage[usenames,dvipsnames]{color}
\usepackage[colorlinks=true,citecolor=blue,linkcolor=magenta]{hyperref}
\usepackage{ulem}
\usepackage{lmodern}

\newcommand{\figpath}{.}

\newcommand{\Tr}{\mathrm{Tr}}

\newcommand{\ket}[1]{\vert{ #1 }\rangle}

\begin{document}

\title{Noise threshold and resource cost of fault-tolerant quantum computing with Majorana fermions in hybrid systems}

\author{Ying Li}

\affiliation{Department of Materials, University of Oxford, Parks Road, Oxford OX1 3PH, United Kingdom}

\date{\today}

\begin{abstract}
Fault-tolerant quantum computing in systems composed of both Majorana fermions and topologically unprotected quantum systems, e.g.~superconducting circuits or quantum dots, is studied in this paper. Errors caused by topologically unprotected quantum systems need to be corrected with error correction schemes, for instance, the surface code. We find that the error-correction performance of such a hybrid topological quantum computer is not superior to a normal quantum computer unless the topological charge of Majorana fermions is insusceptible to noise. If errors changing the topological charge are rare, the fault-tolerance threshold is much higher than the threshold of a normal quantum computer, and a surface-code logical qubit could be encoded in only tens of topological qubits instead of about a thousand normal qubits.
\end{abstract}

\maketitle

\section{Introduction}

Anyons are quasiparticle excitations in topologically ordered phases~\cite{Wen1990} of matter that exhibit exchange statistics different from fermions and bosons~\cite{Wilczek1990}. Topological quantum computing is an exotic but feasible approach for realising quantum information processing by encoding qubits in states defined by anyons~\cite{Kitaev2003, Freedman2003, Nayak2008}. These states are usually topologically protected, i.e.~immune to local perturbations and protected from unfavourable excitations by an energy gap. Quantum gates can be implemented by braiding non-Abelian anyons, which tolerate small perturbations to braiding operations. Majorana fermions (MFs) are non-Abelian anyons which could emerge in the quantum Hall state at filling fraction $\nu = 5/2$~\cite{Moore1991, Read2000}, one-dimensional semiconducting wires coupled to superconductors~\cite{Kitaev2001, Lutchyn2010, Oreg2010} and some other materials in non-Abelian topological phases~\cite{Fu2008, Linder2010, Sau2010, Alicea2010, Sato2009, Lee2009, Ghosh2010, Qi2010}. Evidence of MFs has been observed in recent experiments~\cite{Mourik2012, Mebrahtu2013, NadjPerge2014, Xu2015, Haim2015, Albrecht2016}. However, universal quantum computing is not supported by braiding MFs. Therefore, some computing operations need to be performed with other mechanisms~\cite{Bravyi2006, Freedman2006}, for example, changing the topology of the system~\cite{Bonderson2010}, anyonic interferometry~\cite{Bonderson2006} or coupling MFs to topologically unprotected quantum systems (TUQSs), e.g.~superconducting circuits~\cite{Sau2010PRB, Hassler2010, Hassler2011, Jiang2011, Hyart2013, Schmidt2013, Zhang2013} or quantum dots~\cite{Flensberg2011, Bonderson2011, Leijnse2011}.

\begin{figure}[tbp]
\centering
\includegraphics[width=1\linewidth]{\figpath 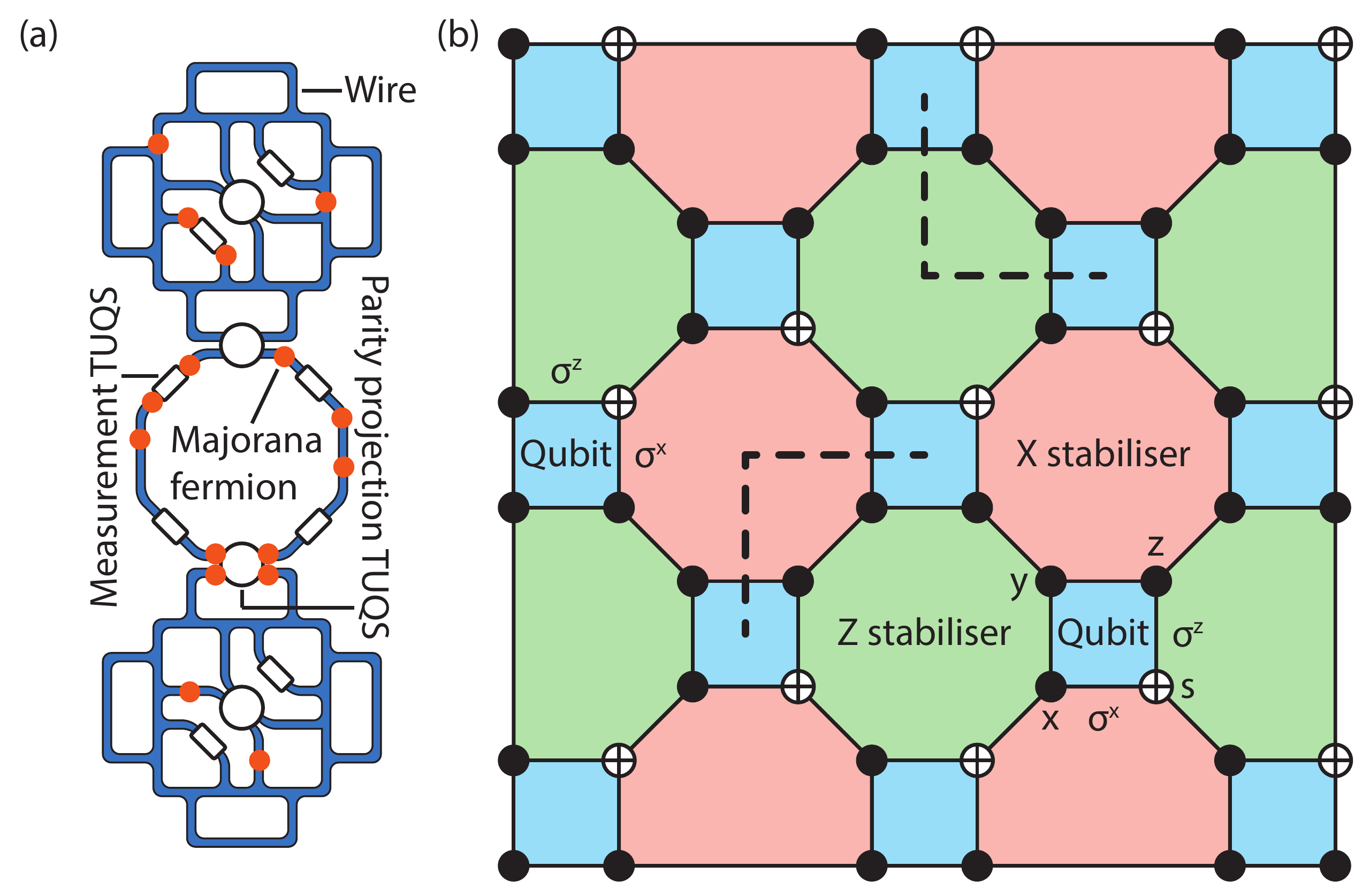}
\caption{
(a) A hybrid platform of implementing the surface code of MF qubits. Measurement TUQSs and PP TUQSs are coupled to one-dimensional semiconducting wires. MFs are created and moved on the wire network by adjusting gate voltages~\cite{Alicea2011}. Measurement and PP operations are performed by coupling MFs to corresponding TUQSs. See Appendix~\ref{app:network} for details of implementing the surface code on such a platform. (b) The surface code of MF qubits. Each blue square denotes a qubit encoded in four MFs $\{ s,x,y,z \}$: the crossed circle denotes $s$, and solid circles respectively denote $x$, $y$ and $z$ in a clockwise direction. Each red (green) octagon corresponds to an X (Z) stabiliser, which is the product of Pauli operators $\sigma^x$ ($\sigma^z$) of four surrounding qubits, i.e.~the stabiliser equals the product of the fermion parity of eight MFs on the octagon and fermion parities of two qubits (marked with dashed lines, see Appendix~\ref{app:StaMea}). Measuring stabilisers needs ancillary MFs, which are not shown in this figure.
}
\label{fig:wire_code}
\end{figure}

In this paper, we consider the hybrid platform, in which TUQSs are used to complete the universality of the quantum computing. We consider a quantum computer composed of three types of elements [Fig.~\ref{fig:wire_code}(a)]: i) topological materials carrying MFs and providing braiding operations, e.g.~one-dimensional semiconducting wires~\cite{Alicea2011}, ii) TUQSs for measuring the topological charge (TC) of two MFs ({\it Measurement})~\cite{Sau2010PRB, Hassler2010, Hassler2011, Hyart2013, Flensberg2011}, and iii) TUQSs for measuring TC of four MFs [{\it Parity projection} (PP)]~\cite{Hyart2013}. We also need TUQSs for preparing magic states~\cite{Sau2010PRB, Hassler2011, Schmidt2013, Zhang2013}, which are not shown in the figure. Compared with a fully topologically protected quantum computer, the drawback of the hybrid platform is obvious. Operations performed with TUQSs are not fault-tolerant themselves, therefore errors caused by TUQSs need to be corrected with error correction codes~\cite{Bravyi2010}.

We find that the performance of the error correction strongly depends on how good TC of MFs is preserved. The fermion parity, i.e.~TC of MFs, could be changed via tunnelling couplings to fermionic baths in the environment~\cite{Goldstein2011, Trauzettel2012, Cheng2012, Loss2012, Campbell2015}, which can be suppressed by avoiding couplings to gapless fermionic baths~\cite{Trauzettel2012}, maintaining the temperature much lower than the energy gap~\cite{Cheng2012} and minimising low-temperature quasiparticle poisoning in the superconductor~\cite{Loss2012}. We will show that a hybrid topological quantum computer is superior to a normal quantum computer if charge errors are rare, i.e. the rate of errors changing TC is much lower than the rate of errors that conserve TC.

In error correction, the infidelity of quantum computing can be suppressed to an arbitrarily low level if the rate of errors per operation is lower than a threshold. In this paper, we propose a protocol of realising the surface code~\cite{Dennis2002, Fowler2009, Terhal2012, Vijay2015, Landau2016} with operations on up to four MFs~\cite{Bravyi2002} and an efficient distillation circuit for assisting the error correction. The surface code only requires entangling gates between nearest neighbouring qubits, and its noise threshold is among the highest recorded. The distillation circuit can reduce errors if most of the errors are charge-conserving~\cite{Bravyi2006}. We find that if populations of charge errors and charge-conserving errors are similar, the threshold is $\sim 0.85\%$, which is comparable to the threshold of a normal-qubit quantum computer~\cite{Fowler2009, Wang2011}, e.g. based on superconducting qubits; If charge errors are rare, the threshold could be as high as $50\%$.

Besides the noise threshold, the overhead cost (number of qubits) for encoding a surface-code logical qubit is another important measure of the error-correction performance. We focus on the case that only one error occurs per thousand topologically unprotected operations, i.e.~the error rate is $\sim 0.1\%$. This error rate has been demonstrated with trapped ions~\cite{Lucas, Wineland} and is also feasible in other quantum systems, e.g. superconducting qubits~\cite{SCQ}. If populations of charge errors and charge-conserving errors are similar, the cost for encoding a logical qubit in the hybrid platform is similar to the cost in a normal-qubit quantum computer~\cite{Fowler2012}; If charge errors are rare, the cost is significantly reduced, i.e.~a logical qubit can be encoded in only tens of Majorana qubits, which usually requires about a thousand normal qubits.

\section{Majorana fermions}

MFs are described by Hermitian operators $\{ c_i \}$ satisfying $\{ c_i , c_j \} = 2\delta_{i,j}$. Each pair of MFs corresponds to a fermionic mode $f = (c_i + ic_j)/2$, where $f$ is the annihilation operator of the fermionic mode. TC of $2n$ MFs is determined by the fermion parity $Q = (-i)^n c_1 c_2 \cdots c_{2n}$~\cite{Bravyi2006, footnote1}, which has two eigenvalues $+1$ and $-1$ corresponding to the vacuum $\mathbf{1}$ and particle $\psi$, respectively.

\section{Qubits}

Each qubit is encoded in four MFs $\{ s,x,y,z \}$ [see Fig.~\ref{fig:wire_code}(b) and Appendix~\ref{app:qubit}]~\cite{Bravyi2006}. Pauli operators of the qubit are $\sigma^x = isx$, $\sigma^y = ixz$ and $\sigma^z = isz$. Four MFs correspond to two fermionic modes and have four different states. Qubit states are in a two-dimensional subspace spanned by two degenerate eigenstates of the fermion parity $sxyz$. Actually, MF $y$ is redundant  for encoding a qubit (see Appendix~\ref{app:qubit}) but a helpful ancillary, as we will show later~\cite{Georgiev2006}.

\section{MF-based quantum computing}

Universal quantum computing can be implemented with operations (see Appendix~\ref{app:MFQC})~\cite{Bravyi2002, Bravyi2006}: i) exchange gate $R_{c_1c_2} = (\openone+c_1c_2)/\sqrt{2}$; ii) creation of two MFs in the eigenstate of $ic_1c_2$; iii) measurement of $ic_1c_2$; iv) non-destructive measurement of $c_1c_2c_3c_4$, i.e.~PP; and v) preparation of the magic state $\ket{\text{A}} = (\ket{0} + e^{i\pi/4}\ket{1})/\sqrt{2}$~\cite{Bravyi2005}. On the hybrid platform, MFs are created on topological materials, exchange gates are performed by braiding MFs, and other operations are performed using corresponding TUQSs.

\section{Surface code}

A logical qubit is encoded in a two-dimensional qubit array [Fig.~\ref{fig:wire_code}(b)]~\cite{Dennis2002, Fowler2009}. The code subspace is spanned by common eigenstates of stabilisers. There are two species of stabilisers: an X (Z) stabiliser is the product of Pauli operators $\sigma^x$ ($\sigma^z$) of four neighbouring qubits [see Fig.~\ref{fig:wire_code}(b) and Appendix~\ref{app:StaMea}]. Errors are detected by repeatedly measuring stabilisers. Universal quantum computing can be fault-tolerantly implemented on logical qubits~\cite{Fowler2009, Horsman2012} by using fault-tolerant operations and injecting magic states into the surface code~\cite{Li2015}.

\section{Stabiliser measurements}
\label{sec:StaMea}

Each full round of stabiliser measurements is performed in four steps [Fig.~\ref{fig:wire_code}(b)]: first, TC of each qubit (four MFs on each \textit{blue} square) is measured; second, TC of eight MFs on each \textit{red} octagon is measured; third, TC of each qubit is measured again; finally, TC of each \textit{green} octagon is measured. The value of a stabiliser equals the product of the corresponding octagon fermion parity and fermion parities of two corresponding qubits (see Appendix~\ref{app:StaMea}). This protocol of stabiliser measurements allows the surface code to tolerate a high error rate when charge errors are rare.

TC of eight MFs is measured using the circuit in Fig.~\ref{fig:circuit}(a) (see Appendix~\ref{app:proof}). Because the total TC of ancillary MFs should be conserved, such a circuit can detect errors that change the ancillary TC.

\begin{figure}[tbp]
\centering
\includegraphics[width=1\linewidth]{\figpath 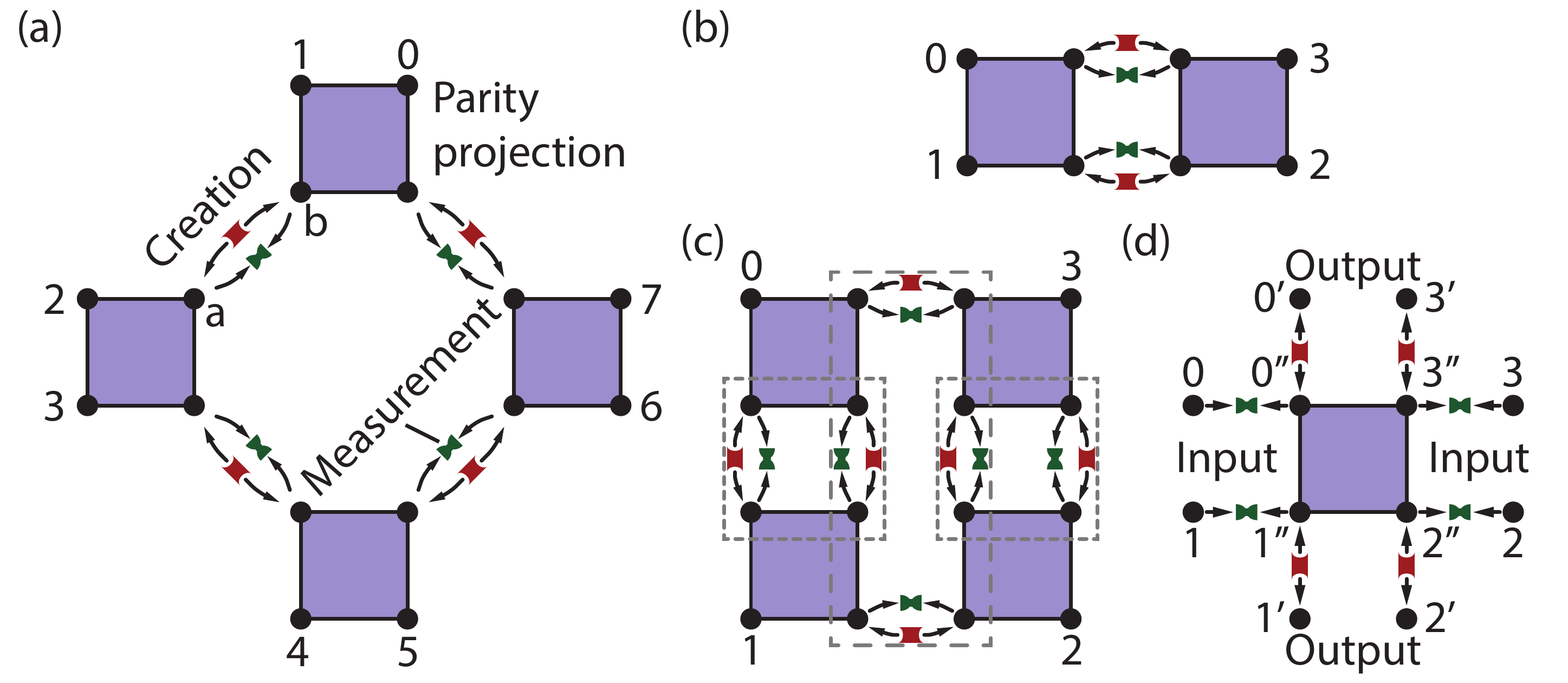}
\caption{
Circuits of (a) the measurement of eight-MF fermion parity $c_0c_1\ldots c_7$, (b) an effective PP with the partial error detection, (c) an effective PP with the full error detection, and (d) an effective PP using an entangled state. In (b)-(d), $c_0c_1c_2c_3$ is measured. MFs without a number are ancillaries. Each purple square denotes a PP, each red pipe with outgoing arrows denotes the creation of a pair of MFs $a$ and $b$ in the eigenstate of $iab$, and each green pipe with incoming arrows denotes the measurement of $iab$. See Appendix~\ref{app:proof} for details.
}
\label{fig:circuit}
\end{figure}

\section{Model of noise}

We use a model of noise to describe the performance of the hybrid system rather than focusing on a specific realisation of the system. Quantum information stored in distant MFs is expected to be immune to local perturbations due to the environment. Most errors occur when MFs are brought together to interact with TUQSs for performing operations: creation, measurement and PP. Therefore, we neglect memory errors, errors in the transfer of MFs and errors in exchange gates~\cite{footnote2}. We model errors in topologically unprotected operations as the following.

A pair of MFs $c_0$ and $c_1$ may be created in the incorrect eigenstate of $ic_0c_1$ with the rate $\epsilon_\text{c}$. The measurement outcome of $ic_0c_1$ may be incorrect with the rate $\epsilon_\text{m}$. For a PP on $\{c_0,c_1,c_2,c_3\}$, when the measurement outcome (the value of $c_0c_1c_2c_3$) is $\mu = \pm 1$, the operation performed on the system is
\begin{eqnarray}
\Pi^{(\mu)} = \mathcal{E}_\text{+} \circ [\pi_{\mu}] + \mathcal{E}_\text{-} \circ [\pi_{-\mu}],
\label{eq:Pimu}
\end{eqnarray}
where
\begin{eqnarray}
\mathcal{E}_\pm = \epsilon_{\pm,0}[\openone] + \frac{\epsilon_{\pm,1}}{4}\sum_{i=0}^3[c_i]
+ \frac{\epsilon_{\pm,2}}{6} \sum_{i=0}^2\sum_{j=i+1}^3[c_ic_j].
\label{eq:Epm}
\end{eqnarray}
Here, the superoperator $[U]\rho  = U\rho U^\dag$, and the projector $\pi_{\mu} = (\openone + \mu c_0c_1c_2c_3)/2$. $\epsilon_{+,0}$ is the fidelity, $\epsilon_{-,0}$, $\epsilon_{\pm,1}$ and $\epsilon_{\pm,2}$ are rates of corresponding errors, and $\sum_{l=0}^2 (\epsilon_{+,l}+\epsilon_{-,l}) = 1$. $\mathcal{E}_\text{+} \circ [\pi_{\mu}]$ and $\mathcal{E}_\text{-} \circ [\pi_{-\mu}]$ respectively denote correct and incorrect measurement outcomes. In the scenario of implementing our surface-code protocol, any noisy PP, as long as it only affects four relevant MFs, can be converted into an operation in the form of Eqs.~(\ref{eq:Pimu})~and~(\ref{eq:Epm}) using some redundant exchange gates (see Appendix~\ref{app:model}).

Errors are classified into two categories: errors that change TC (i.e. $[c_i]$) and errors that are equivalent to erroneous braidings (i.e.~$[c_ic_j]$). We classify creation and measurement errors into the first category (see Appendix~\ref{app:tunnelling}), which result in incorrect records of TC. For simplification, we assume $\epsilon_\text{c} = \epsilon_\text{m} = \epsilon_{+,1} = \epsilon_{-,0} = \epsilon_{-,1} = \epsilon_{-,2} = p_\text{F}$ and $\epsilon_{+,2} = p_\text{B}$ corresponding to charge (fermion-parity) errors and charge-conserving (braiding) errors, respectively. We would like to remark that charge-conserving errors only occur in PPs, all errors in creation and measurement operations are charge errors, and these two kinds of errors can be corrected in different manners.

\section{Error correction and thresholds}

Errors that could occur on a qubit are $\{[s],[x],[y],[z]\}$ and their combinations. Because the qubit is either encoded in the subspace $sxyz = 1$ or $sxyz = -1$, the error $[sxyz]$ has no effect on the qubit state, i.e.~can be ignored. Then, we have $[s] = [y][\sigma^y]$, $[x] = [y][\sigma^z]$ and $[z] = [y][\sigma^x]$. Therefore, errors are all equivalent to combinations of Pauli errors and the $[y]$ error. Because MF $y$ is redundant, it is not necessary to correct the $[y]$ error,  which changes TC of the qubit.

Pauli errors are corrected with the surface code (see Appendix~\ref{app:scheme})~\cite{Fowler2009}. Without charge errors, i.e.~$p_\text{F} = 0$, outcomes of stabiliser measurements are always correct using our protocol of stabiliser measurements. In this limit, the threshold of independent Pauli errors is $\sim 10\%$~\cite{Dennis2002}. In our protocol, $[\sigma^x]$ and $[\sigma^z]$ occur with the same rate $\sim 4p_\text{B}$ when $p_\text{F} = 0$, which results in the threshold $p_\text{B} = 2.7\%$. If $p_\text{F} = p_\text{B}$, the threshold is $p_\text{B} = 0.15\%$, i.e.~the fidelity of PPs is $99.25\%$.

Although it is not necessary to correct qubit-charge errors in the form $[y]$, one can obtain more information about Pauli errors by detecting qubit-charge errors, which utilises correlations between errors. In the case $p_\text{F} = 0$, when a $[y]$ error occurs, an $[a]$ error and some Pauli errors always occur simultaneously (see Appendix~\ref{app:scheme}). Here, $a$ is an ancillary MF for measuring TC of an octagon neighbouring to the qubit of $y$. The $[y]$ error can be detected by the qubit-charge measurement, and the $[a]$ error can be detected in the octagon-charge measurement (see Appendix~\ref{app:scheme}). Once an $[ay]$ error is detected, corresponding Pauli errors are corrected. With this procedure, the rate of Pauli errors is reduced to $\sim 8p_\text{B}/3$ when $p_\text{F} = 0$, and the threshold is increased to $p_\text{B} = 4.1\%$. When $p_\text{F} > 0$, it is still helpful to correct Pauli errors according to detected $[ay]$ errors despite the fact that $[y]$, $[a]$ and these Pauli errors may occur independently. By detecting $[y]$ errors and $[a]$ errors, the threshold is increased to $p_\text{B} = 0.17\%$ when $p_\text{F} = p_\text{B}$, i.e.~the fidelity of PPs is $99.15\%$.

\begin{figure}[tbp]
\centering
\includegraphics[width=1\linewidth]{\figpath 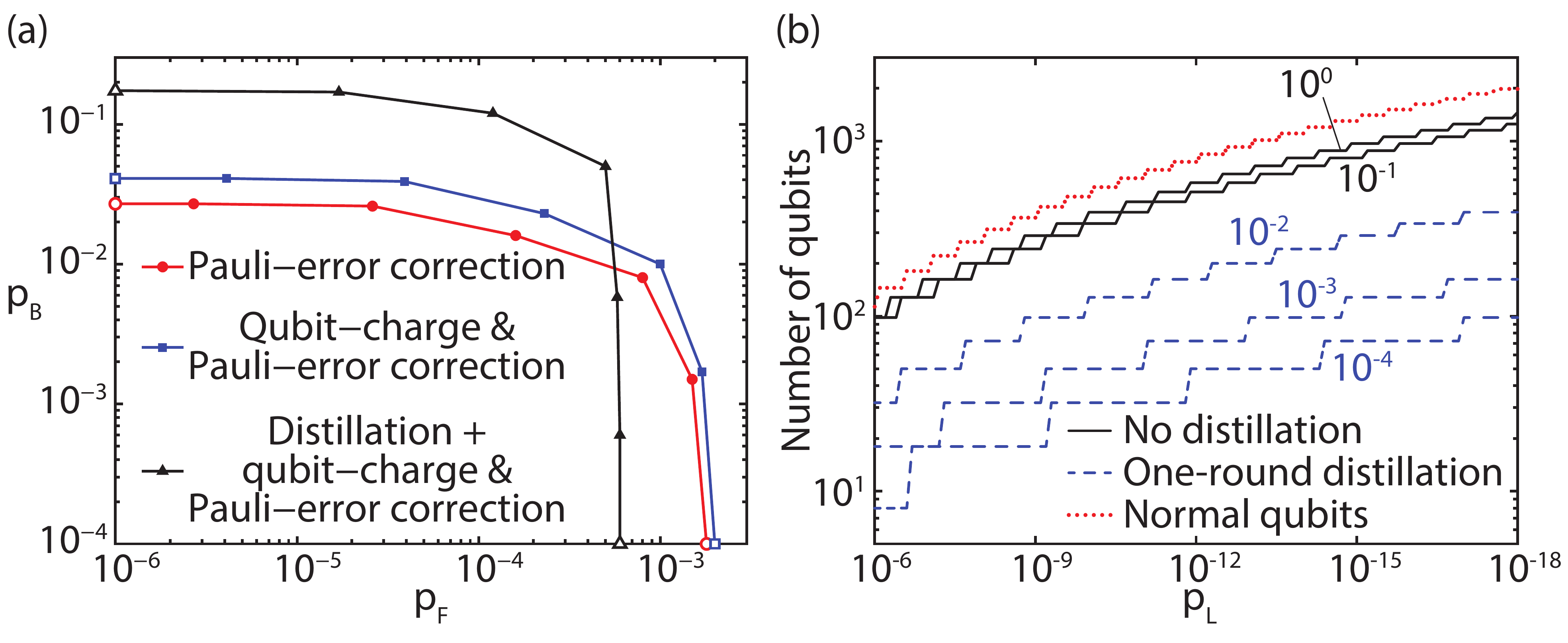}
\caption{
(a) Thresholds of the MF-based fault-tolerant quantum computing. On each curve, from left to right, the ratio $p_\text{F}/p_\text{B}$ of marks are $0,10^{-4},10^{-3},10^{-2},10^{-1},1,\infty$, respectively, i.e.~empty marks on the vertical (horizontal) axis correspond to $p_\text{F} = 0$ ($p_\text{B} = 0$). Assisted by detecting qubit-charge errors, the threshold is higher than only detecting Pauli errors. When $p_\text{F} \ll p_\text{B}$, the threshold can be improved by using distilled PPs, where the distillation is only performed for one round. (b) Number of qubits for encoding a logical qubit to achieve the logical-qubit error rate $p_\text{L}$ (per round of stabiliser measurements). The fidelity of PPs is $\sim 99.9\%$, i.e.~$p_\text{B} = 0.02\%$ when $p_\text{F}/p_\text{B} = 1$, and $p_\text{B} = 0.1\%$ when $p_\text{F}/p_\text{B} = 10^{-1},10^{-2},10^{-3},10^{-4}$. For normal qubits, stabiliser measurements are implemented using CNOT gates with the fidelity $99.9\%$~\cite{Fowler2012}. From top to bottom, solid lines respectively correspond to $p_\text{F}/p_\text{B} = 1,10^{-1}$, and dashed lines respectively correspond to $p_\text{F}/p_\text{B} = 10^{-2},10^{-3},10^{-4}$. Data are obtained numerically. See Appendix~\ref{app:data} for more data and details.
}
\label{fig:plots}
\end{figure}

Thresholds for a variety of $p_\text{F} / p_\text{B}$ are shown in Fig.~\ref{fig:plots}(a).

\section{Distillation}

If charge errors are rare, i.e.~$p_\text{F} \ll p_\text{B}$, distillation circuits can eliminate charge-conserving errors, which only occur in PPs. In the previous protocol~\cite{Bravyi2006}, errors in maximally entangled two-qubit states are reduced in the distillation if the total TC of two qubits is accurately determined and the error rate is lower than $\sim 38.4\%$. We propose a distillation protocol based on PPs: without charge errors, i.e.~$p_\text{F} = 0$, the rate of charge-conserving errors is reduced in the distillation if $p_\text{B} < 50\%$; if $p_\text{B} \ll 50\%$, the error rate is reduced from $p_\text{B}$ to $\sim p_\text{B}^2$ in each round of the distillation.

Before giving our protocol, we introduce three types of effective PPs.

In the effective PP in Fig.~\ref{fig:circuit}(b), similar to the circuit in Fig.~\ref{fig:circuit}(a), the total ancillary TC should be conserved. Therefore, charge-conserving errors exchanging TC between $\{c_0,c_1,c_2,c_3\}$ and ancillary MFs~\cite{footnote3}, i.e.~changing the ancillary TC, can be detected.

All types of charge-conserving errors can be detected in the effective PP in Fig.~\ref{fig:circuit}(c). This circuit is the second generation of the circuit in Fig.~\ref{fig:circuit}(b), i.e.~each raw PP is replaced with a first-generation effective PP. TC of ancillary MFs in each dashed-line box in Fig.~\ref{fig:circuit}(c) should be conserved. Therefore, any charge-conserving errors in four raw PPs can be detected. If ancillary TCs are not changed by chance, the error rate of this effective PP is
\begin{eqnarray}
\frac{ \frac{8}{9}p_\text{B}^2 \left( 1-\frac{5}{3}p_\text{B}+\frac{7}{9}p_\text{B}^2 \right) }
{ 1-4p_\text{B} + \frac{32}{9}p_\text{B}^2 \left( 2-\frac{5}{3}p_\text{B}+\frac{5}{9}p_\text{B}^2 \right) }
\underset{p_\text{B} \ll 1}\simeq \frac{8}{9}p_\text{B}^2,
\end{eqnarray}
where we have assumed that $p_\text{F} = 0$, and $p_\text{B}$ is the error rate of raw PPs. Here, the denominator of LHS is the probability that ancillary TCs are unchanged. One can find that LHS is smaller than $p_\text{B}$ if $p_\text{B} < 50\%$.

In the effective PP in Fig.~\ref{fig:circuit}(d), a raw PP is used to create a two-qubit entangled state on MFs $\{c_{0'},c_{1'},c_{2'},c_{3'}\}$ and $\{c_{0''},c_{1''},c_{2''},c_{3''}\}$, and this entangled state is consumed to perform an effective PP on MFs $\{c_0,c_1,c_2,c_3\}$ as proposed in Ref.~\cite{Bravyi2006}. In this effective PP, the state of $\{c_0,c_1,c_2,c_3\}$ is transferred to $\{c_{0'},c_{1'},c_{2'},c_{3'}\}$ and $c_0c_1c_2c_3$ is measured simultaneously.

In our distillation protocol, we replace the raw PP in Fig.~\ref{fig:circuit}(d) with the error-detection PP in Fig.~\ref{fig:circuit}(c). The error-detection PP can only succeed probabilistically, i.e.~it fails when errors are detected. To select a successful effective PP, we use the error-detection PP to create a two-qubit entangled state and use the state, only if no error is detected, to carry out the final PP.

Without charge errors, one can eliminate all errors by iterating distillation circuits to achieve fault-tolerant quantum computing. However, if $p_\text{F}$ is finite, the error correction code is necessary because charge errors accumulate in the distillation and need to be corrected in a different manner. In general, when $p_\text{F} \ll p_\text{B}$, the distillation can improve the threshold of the error correction. Using PPs after one round of the full-error distillation, the threshold is significantly increased when $p_\text{F} \leq 10^{-2} p_\text{B}$ [Fig.~\ref{fig:plots}(a)].

\section{Resource cost}

In Fig.~\ref{fig:plots}(b), we show the cost of qubits for encoding a logical qubit as a function of the logical-qubit error rate. We focus on the PP error rate $\sim 0.1\%$, which is realistic and lower than the threshold for any ratio $p_\text{F}/p_\text{B}$. The rate of logical-qubit errors scales as $p_\text{L} \simeq \exp(-\kappa d - \nu \log d - \eta)$~\cite{Raussendorf2007NJP, Fowler2013a, Watson2014}, where $d$ is the size of the lattice [the number of qubits on the side, which is $3$ in Fig.~\ref{fig:wire_code}(b)], and parameters $\kappa$, $\nu$ and $\eta$ are found numerically (see Appendix~\ref{app:data}). When the required logical-qubit error rate is lower, more qubits are needed for encoding a logical qubit. As a comparison, the cost of normal qubits~\cite{Fowler2012} is also plotted in Fig.~\ref{fig:plots}(b). If $p_\text{F} = p_\text{B}$, costs of Majorana qubits and normal qubits are similar. If $p_\text{F}$ is much lower than $p_\text{B}$, i.e.~$p_\text{F} \lesssim 10^{-2} p_\text{B}$, the qubit cost can be significantly reduced by using the distillation. If $p_\text{F} \lesssim 10^{-4} p_\text{B}$, a logical qubit can be encoded in about fifty Majorana qubits to achieve the logical-qubit error rate $5\times 10^{-15}$, which requires more than a thousand normal qubits.

The time cost is discussed in Appendix~\ref{app:time}.

\section{Conclusions}

We have studied the topological quantum computing based on MFs in hybrid systems. So far, we have assumed that memory errors are negligible. Memory errors occur when MFs are decoupled from TUQSs and far apart from each other, in which case the charge exchange between MFs (charge-conserving errors) is suppressed, and most memory errors are charge errors. If the rate of memory errors (errors occurring between two operations) is much lower than the rate of charge errors in operations, our conclusions will be the same. It is similar for errors in transfer and exchange operations. A noise threshold $0.85\%$ is obtained for entangling operations performed with interactions between MFs and TUQSs. The threshold is much higher (up to $50\%$), if TC of MFs is preserved. The overhead cost in encoding logical qubits is also studied. Given the feasible error rate $\sim 0.1\%$, a high quality logical qubit can be realised with fewer than a hundred Majorana qubits. Compared with normal qubits, the qubit overhead has been reduced by a factor $\gtrsim 10$, which shows potential advantages of the quantum computing in hybrid MF systems. However, this overhead reduction requires that the rate of charge errors is at least two orders of magnitude lower than the rate of charge-conserving errors, i.e.~hybrid MF computers are fault-tolerant if charge-tunnelling noise is in the sub-threshold regime but have superiority over normal-qubit computers only if charge-tunnelling noise is suppressed to a much lower level.

\begin{acknowledgments}
This work was supported by the EPSRC platform grant `Molecular Quantum Devices' (EP/J015067/1), and the EPSRC National Quantum Technology Hub in Networked Quantum Information Processing. We would like to thank Simon C. Benjamin, Leong Chuan Kwek and David Herrera-Mart\'{i} for helpful discussions. The authors would like to acknowledge the use of the University of Oxford Advanced Research Computing (ARC) facility in carrying out this work. http://dx.doi.org/10.5281/zenodo.22558.
\end{acknowledgments}

%%%%%%%%%%%%%%%%%%%%%%%%%%%%%
%\newpage
%\newpage

%\widetext

\appendix

\section{Qubits}
\label{app:qubit}

Each qubit is encoded in four MFs $\{s,x,y,z\}$~\cite{Bravyi2006}. Pauli operators of the qubit are $\sigma^x = isx$, $\sigma^y = ixz$ and $\sigma^z = isz$. These four MFs correspond to two fermionic modes described by annihilation operators $f_{sz} = (s+iz)/2$ and $f_{xy} = (x+iy)/2$, respectively. Hence, the Hilbert space is four-dimensional and spanned by
$$\{ \ket{0_{sz},0_{xy}},\ket{0_{sz},1_{xy}},\ket{1_{sz},0_{xy}},\ket{1_{sz},1_{xy}} \},$$
where $n_{sz} = 0_{sz},1_{sz}$ and $n_{xy} = 0_{xy},1_{xy}$ are respectively particle numbers of fermionic modes $f_{sz}$ and $f_{xy}$, i.e.~$\ket{n_{sz},n_{xy}} = f_{sz}^{\dag n_{sz}}f_{xy}^{\dag n_{xy}}\ket{0_{sz},0_{xy}}$. This Hilbert space contains two subspaces respectively corresponding to two eigenvalues of the operator (TC of the qubit) $sxyz = -(2f_{sz}^\dag f_{sz} - 1)(2f_{xy}^\dag f_{xy} - 1)$. In the subspace with $sxyz = 1$, qubit states are
\begin{eqnarray}
\ket{0} &\equiv & \ket{1_{sz},0_{xy}}, \\
\ket{1} &\equiv & -i\ket{0_{sz},1_{xy}}.
\end{eqnarray}
In the subspace with $sxyz = -1$, qubit states are
\begin{eqnarray}
\ket{0} &\equiv & \ket{1_{sz},1_{xy}}, \\
\ket{1} &\equiv & -i\ket{0_{sz},0_{xy}}.
\end{eqnarray}
Pauli operators can be rewritten as
\begin{eqnarray}
\sigma^x &=& i(f_{sz}+f_{sz}^\dag)(f_{xy}+f_{xy}^\dag), \\
\sigma^y &=& -(f_{sz}-f_{sz}^\dag)(f_{xy}+f_{xy}^\dag), \\
\sigma^z &=& (2f_{sz}^\dag f_{sz} - 1).
\end{eqnarray}

MF $y$ is redundant. Two qubits can be encoded in only six MFs $\{s_1,x_1,z_1\}$ and $\{s_2,x_2,z_2\}$~\cite{Georgiev2006}. When the fermion parity $is_1x_1z_1s_2x_2z_2$ is determined, the corresponding subspace is four-dimensional, hence two qubits can be encoded. Denser encoding is possible~\cite{Freedman2006}. At most $n-1$ qubits can be encoded in $2n$ MFs.

We would like to choose the four-MF encoding, because Pauli operators of each qubit are defined on its own MFs, which is convenient for implementing the surface code. MF $y$ is a helpful ancillary: with MF $y$, eight-MF TC measurements of the same species (there are two species of stabilisers: X stabilisers and Z stabilisers) can be performed in parallel; by detecting TC of each qubit, more information about errors can be obtained, which helps the error correction.

\begin{figure*}[tbp]
\centering
\includegraphics[width=0.8\linewidth]{\figpath 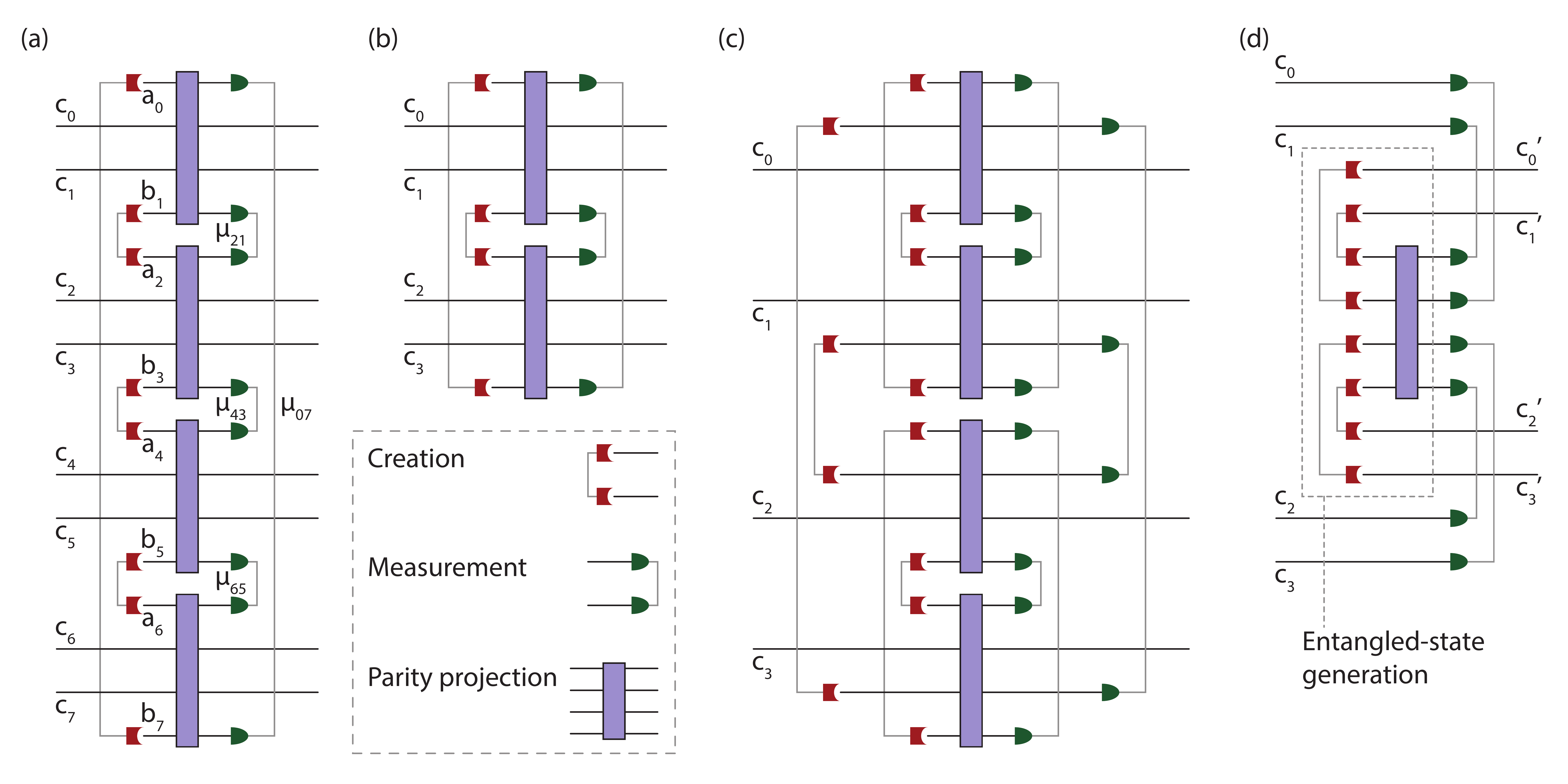}
\caption{
Circuits of (a) the measurement of eight-MF fermion parity $c_0c_1\ldots c_7$, (b) an effective PP with the partial error detection, (c) an effective PP with the full error detection, and (d) an effective PP using an entangled state. In (b)-(d), $c_0c_1c_2c_3$ is measured. These circuits are the same as circuits in Fig.~\ref{fig:circuit}. Each horizontal line denotes a MF. MFs $a$ and $b$ in (a) and MFs without labels in other figures are ancillary MFs.
}
\label{fig:circuit_t}
\end{figure*}

\begin{table}[tbp]
\begin{center}
\begin{tabular}{|c|c|c|c|c|}
\hline
$\mu_{07}$  & $\mu_{21}$ & $\mu_{43}$ & $\mu_{65}$ & $U$ \\ \hline
$1$  & $1$ & $1$ & $1$ & $\openone$ \\ \hline
$-1$  & $-1$ & $1$ & $1$ & $c_0c_1$ \\ \hline
$1$  & $-1$ & $-1$ & $1$ & $c_2c_3$ \\ \hline
$1$  & $1$ & $-1$ & $-1$ & $c_4c_5$ \\ \hline
$-1$  & $1$ & $1$ & $-1$ & $c_6c_7$ \\ \hline
$-1$  & $1$ & $-1$ & $1$ & $c_0c_1c_2c_3$ \\ \hline
$1$  & $-1$ & $1$ & $-1$ & $c_2c_3c_4c_5$ \\ \hline
$-1$  & $-1$ & $-1$ & $-1$ & $c_0c_1c_4c_5$ \\ \hline
\end{tabular}
\end{center}
\caption{
The feedback operation $U$ of the eight-MF TC measurement. We would like to remark that feedback operations $U$ and $UQ_8$ are equivalent to each other. $\mu_{ij}$ is the measurement outcome of $ia_ib_j$ (see Fig.~\ref{fig:circuit_t}).
}
\label{tab} 
\end{table}

\section{Majorana-fermion-based Quantum computing}
\label{app:MFQC}

Operations i--iv provide the full set of Clifford operations: $S = \text{diag}(1,i)$ gate and Hadamard gate can be performed using exchange gates; initialisation and measurement of a qubit can be performed using the creation and measurement of MFs; the two-qubit entangling gate (e.g.~CNOT gate) can be performed using PPs~\cite{Bravyi2002}. Actually, PP, the two-qubit entangling gate and the preparation of a maximally entangled two-qubit (eight-MF) state can simulate one another together with operations i--iii~\cite{Bravyi2006}. Consuming qubits in the magic state $\ket{\text{A}}$, $T = \text{diag}(1,e^{i\pi/4})$ gates can be performed using Clifford operations~\cite{Bravyi2005}.

The qubit state $\ket{0}$ in the subspace $sxyz = \pm 1$ can be prepared by creating a pair of MFs in the eigenstate $isz = 1$, and another pair of MFs in the eigenstate $ixy = \mp 1$. Measurement on the qubit in the $\sigma^z$ basis is equivalent to the measurement of $isz$.

The phase gate is equivalent to an exchange gate,
\begin{eqnarray}
S = \frac{e^{i\pi/4}}{\sqrt{2}} (\openone - i\sigma^z) = e^{i\pi/4} R_{sz}.
\end{eqnarray}
The Hadamard gate can also be achieved using exchange gates,
\begin{eqnarray}
H = \frac{1}{\sqrt{2}} (\sigma^x + \sigma^z) = iR_{sx}^2R_{xz}.
\end{eqnarray}

PP of four MFs can be used to perform parity projections of two qubits, i.e.~non-destructive measurements of $\sigma^x_1 \sigma^x_2 = -s_1x_1s_2x_2$ or $\sigma^z_1 \sigma^z_2 = -s_1z_1s_2z_2$. Using qubit parity projections, one can implement CNOT gates. We assume that the input state of control and target qubits is $\alpha\ket{0+}_\text{ct}+\beta\ket{0-}_\text{ct}+\gamma\ket{1+}_\text{ct}+\delta\ket{1-}_\text{ct}$. We need an ancillary qubit in the state $\ket{+}_\text{a}$, which is obtained by performing a Hadamard gate on a qubit initialised in the state $\ket{0}_\text{a}$. To implement the CNOT gate,
\begin{itemize}
\item[1.] Firstly, the parity projection $\sigma^z_\text{c} \sigma^z_\text{a}$ is performed on the control qubit and the ancillary qubit. Depending on the measurement outcome $\mu_{\sigma^z_\text{c} \sigma^z_\text{a}}$, the gate $(\sigma^x_\text{a})^{\mu_{\sigma^z_\text{c} \sigma^z_\text{a}}}$ is performed.
\item[2.] Secondly, the parity projection $\sigma^x_\text{t} \sigma^x_\text{a}$ is performed on the ancillary qubit and the target qubit. Depending on the measurement outcome $\mu_{\sigma^x_\text{t} \sigma^x_\text{a}}$, the gate $(\sigma^z_\text{c} \sigma^x_\text{a})^{\mu_{\sigma^x_\text{t} \sigma^x_\text{a}}}$ is performed.
\item[3.] Finally, the ancillary qubit is measured in the $\sigma^z_\text{a}$ basis. Depending on the measurement outcome $\mu_{\sigma^z_\text{a}}$, the gate $(\sigma^x_\text{t} \sigma^x_\text{a})^{\mu_{\sigma^z_\text{a}}}$ is performed.
\end{itemize}
Then, the output state is $(\alpha\ket{0+}_\text{ct}+\beta\ket{0-}_\text{ct}+\gamma\ket{1+}_\text{ct}-\delta\ket{1-}_\text{ct})\ket{0}_\text{a}$, i.e.~a CNOT gate has been performed on control and target qubits.

\begin{figure}[tbp]
\centering
\includegraphics[width=0.5\linewidth]{\figpath 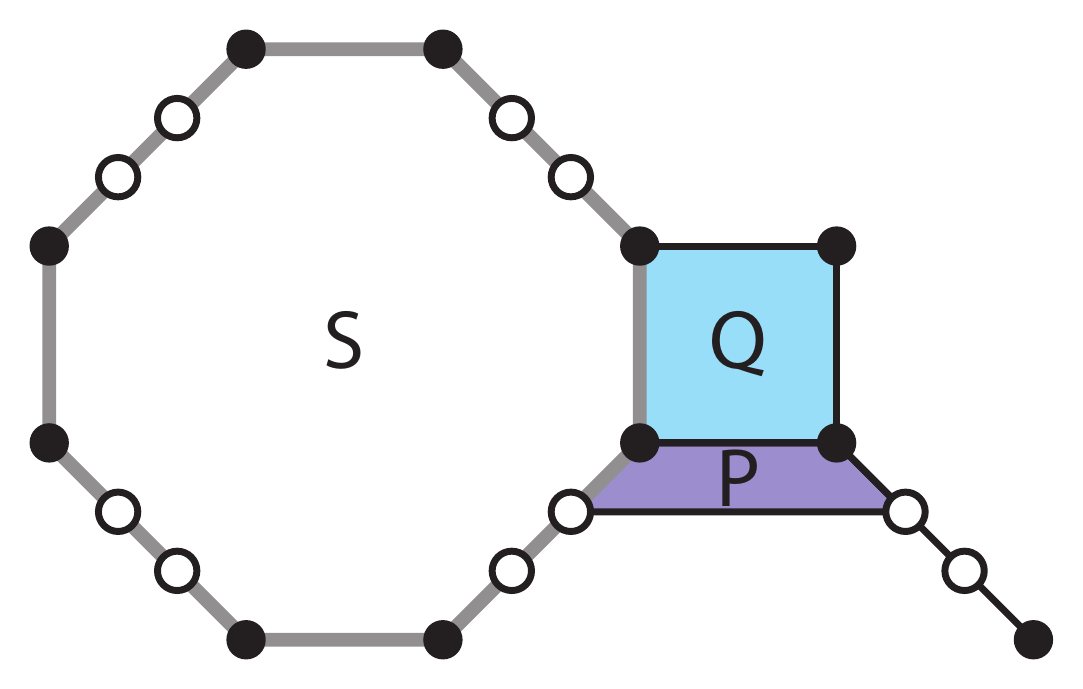}
\caption{
Charges that can remove unwanted cross terms in a noisy PP for measuring TC of an octagon. Filled circles are qubit MFs, and empty circles are ancillary MFs. The purple trapezoid denotes the noisy PP, and the blue square denotes a qubit. Three charges are $P$, $Q$ and $S$, which are TC of four measured MFs, TC of the qubit and TC of all MFs (including ancillary MFs) on a neighbouring octagon (gray bold line).
}
\label{fig:stabilisers}
\end{figure}

\section{Stabiliser measurements}
\label{app:StaMea}

There are two kinds of stabilisers $S_\text{X} = \sigma^x_\text{n} \sigma^x_\text{e} \sigma^x_\text{s} \sigma^x_\text{w} = s_\text{n}x_\text{n} s_\text{e}x_\text{e} s_\text{s}x_\text{s} s_\text{w}x_\text{w}$ and $S_\text{Z} = \sigma^z_\text{n} \sigma^z_\text{e} \sigma^z_\text{s} \sigma^z_\text{w} = s_\text{n}z_\text{n} s_\text{e}z_\text{e} s_\text{s}z_\text{s} s_\text{w}z_\text{w}$, where $\{ \text{n},\text{e},\text{s},\text{w} \}$ are respectively labels of qubits in the north, east, south and west of the corresponding stabiliser. For a red octagon [see Fig.~\ref{fig:wire_code}(b)], TC of eight MFs on the octagon is $Q_\text{R} = s_\text{n}x_\text{n} y_\text{e}z_\text{e} y_\text{s}z_\text{s} s_\text{w}x_\text{w}$. The stabiliser corresponding to the red octagon can be rewritten as $S_\text{X} = Q_\text{R}Q_\text{e}Q_\text{s}$, where $Q_\text{q} = s_\text{q}x_\text{q}y_\text{q}z_\text{q}$ ($\text{q} = \text{n},\text{e},\text{s},\text{w}$) is TC of a qubit. Similarly, for a green octagon [see Fig.~\ref{fig:wire_code}(b)], TC of eight MFs on the octagon is $Q_\text{G} = x_\text{n}y_\text{n} x_\text{e}y_\text{e} s_\text{s}z_\text{s} s_\text{w}z_\text{w}$. The stabiliser corresponding to the green octagon can be rewritten as $S_\text{Z} = Q_\text{G}Q_\text{n}Q_\text{e}$. Therefore, stabilisers can be measured by measuring TCs of qubits and octagons.

\begin{figure}[tbp]
\centering
\includegraphics[width=1\linewidth]{\figpath 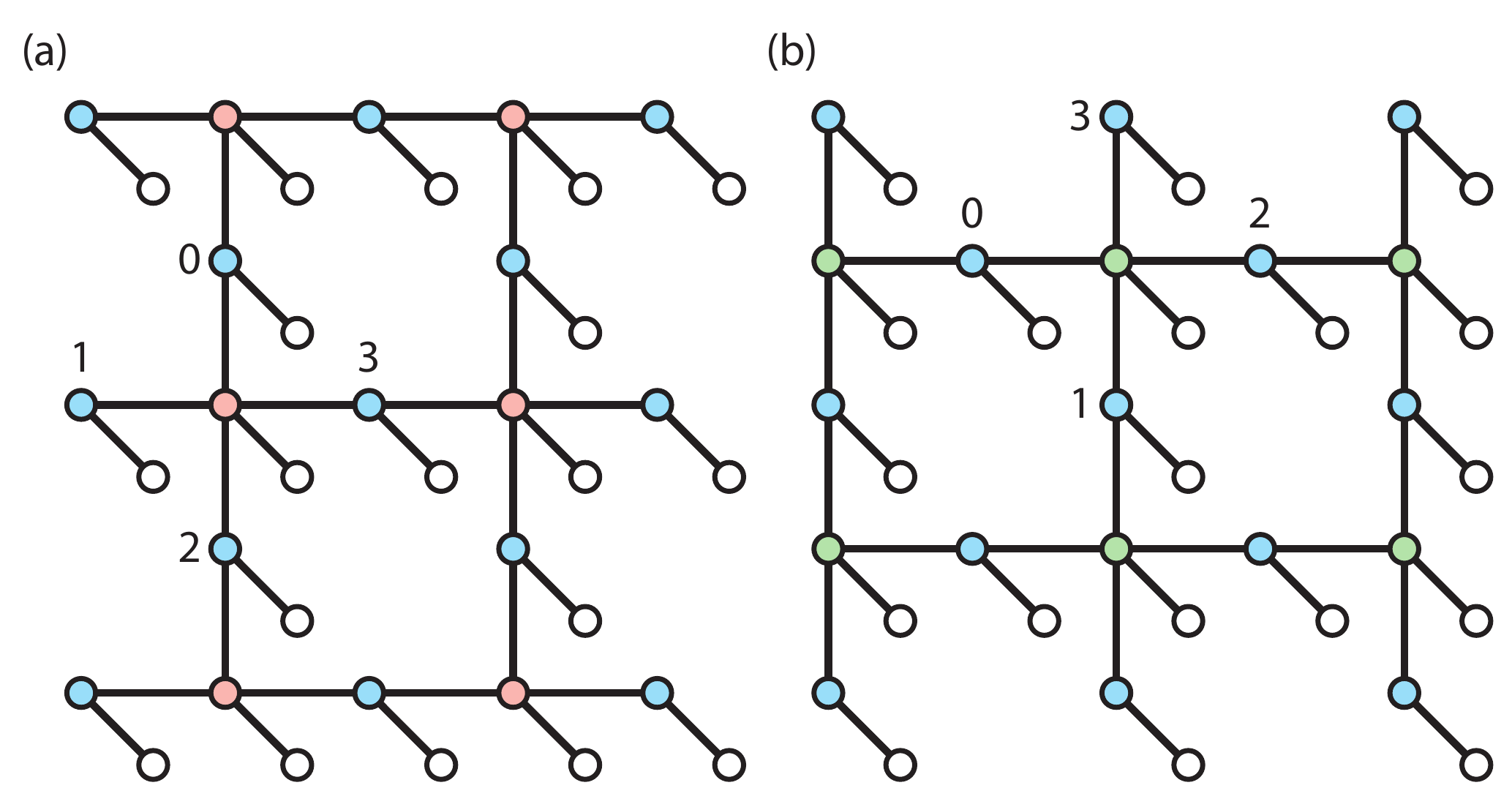}
\caption{
Error-correction lattice of qubit-charge errors.
}
\label{fig:lattice}
\end{figure}

\section{Eight-Majorana-fermion topological charge measurement}
\label{app:proof}

To measure TC of eight MFs $Q_8 = c_0c_1\cdots c_7$ [see Fig.~\ref{fig:circuit}(a) and Fig.~\ref{fig:circuit_t}(a)], eight ancillary MFs are introduced. Each pair of ancillary MFs $a$ and $b$ is initialised in the state $iab = 1$, and $iab$ is measured after PPs. In this circuit, the total TC of ancillary MFs (the product of four pairs of $iab$, i.e.~$Q_\text{A} = a_0b_1a_2b_3a_4b_5a_6b_7 = -1$) is known and conserved, i.e.~commutes with PPs ($Q_{01} = a_0c_0c_1b_1$, $Q_{23} = a_2c_2c_3b_3$, $Q_{45} = a_4c_4c_5b_5$ and $Q_{67} = a_6c_6c_7b_7$). Therefore, the value of $Q_8$ is given by the total charge of all sixteen MFs (product of outcomes of four PPs) and the total charge of ancillaries, i.e.~$Q_8 = Q_{01}Q_{23}Q_{45}Q_{67}Q_\text{A}$. Because the total ancillary charge is conserved, such a circuit can detect errors: if the total ancillary charge is changed due to errors, the initialisation and measurement of ancillaries should give different values of the total ancillary charge, i.e.~$Q_\text{A} = 1$ when ancillaries are measured.

The Hilbert space of eight MFs is $2^4$-dimensional and equivalent to four qubits. The state of each qubit is fully described by Pauli operators $\{ \sigma^z, \sigma^x \}$. Pauli operators of these four qubits can be chosen as $\{ c_0c_1, c_1c_2 \}$, $\{ c_0c_1c_2c_3, c_3c_4 \}$, $\{ c_0c_1c_2c_3c_4c_5, c_5c_6 \}$, and $\{ c_0c_1c_2c_3c_4c_5c_6c_7, c_7 \}$. We would like to remark that the fourth qubit is not physical ($c_7$ is a Hermitian operator but cannot be physically measured). In the circuit of the eight-MF TC measurement [Fig.~\ref{fig:circuit}(a) and Fig.~\ref{fig:circuit_t}(a)], all these Pauli operators except $c_7$ are conserved up to some feedback operations depending on measurement outcomes of ancillary TCs (see Table~\ref{tab}). Therefore, the state of the first three qubits are not changed by the circuit, and only the fourth qubit (the operator $Q_8 = c_0c_1c_2c_3c_4c_5c_6c_7$) is measured. Other circuits in Fig.~\ref{fig:circuit} and Fig.~\ref{fig:circuit_t} can be proved similarly.

\begin{figure*}[tbp]
\centering
\includegraphics[width=1\linewidth]{\figpath 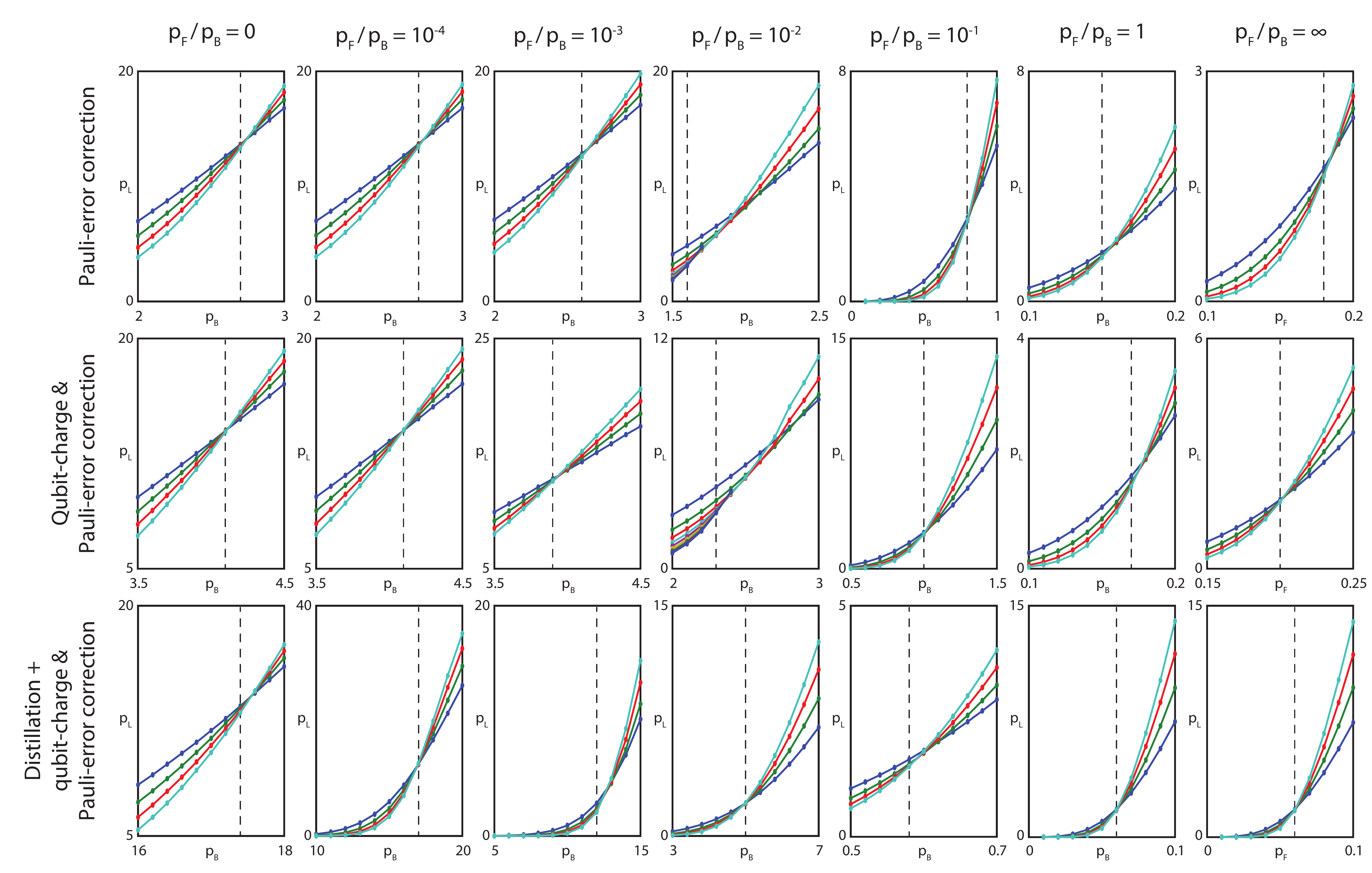}
\caption{
Rate of logical errors per round of stabiliser measurements $p_\text{L}$ (percentage) as a function of the error rate $p_\text{B}$ (percentage). In each figure, the estimated threshold is marked with the vertical dashed line. When $p_\text{B}$ is lower than the threshold, $p_\text{L}$ decreases with the lattice size $d$. On the left side of the vertical line, from top to bottom, curves correspond to $d = 5,7,9,11$, respectively. When $p_\text{F}/p_\text{B} = 10^{-2}$ and the distillation is not used, the finite size effect is significant, so we have also considered larger lattices, i.e.~$d = 13,15,17,19$, to determine the threshold.
}
\label{fig:data}
\end{figure*}

\begin{figure*}[tbp]
\centering
\includegraphics[width=1\linewidth]{\figpath 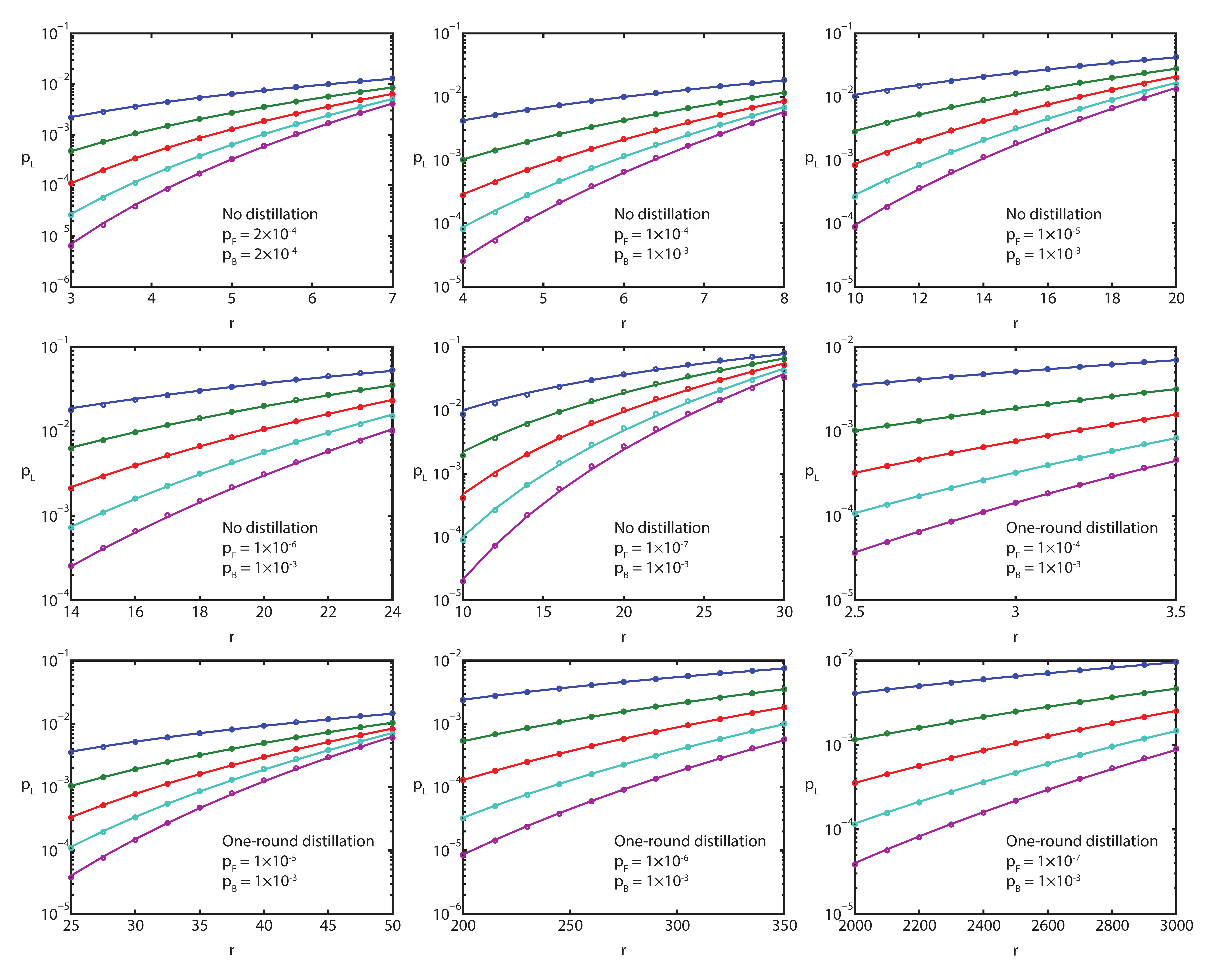}
\caption{
Rate of logical errors per round of stabiliser measurements $p_\text{L}$ as a function of the ratio $r$. In each figure, from top to bottom, curves correspond to $d = 3,5,7,9,11$, respectively. Circles are obtained numerically using the Monte Carlo method, and lines are obtained by fitting circles using Eq.~(\ref{eq:fitting}). Qubit-charge errors are corrected in order to improve the error correction performance.
}
\label{fig:scaling}
\end{figure*}

\section{Universality of the error model}
\label{app:model}

Our error model of noisy PPs is symmetric for two outcomes and four MFs. Because $c_0c_1c_2c_3$ is measured, an error in the form $[c_0c_1c_2c_3]$ has no effect on the state, and errors in the form $[c_ic_jc_k]$ are all equivalent to errors in the form $[c_i]$. The operation $[c_i]$ is not physical for a closed system. Errors $[c_i]$ occur when MFs are coupled with fermionic modes in the environment, i.e.~$([c_i]+[c_j])/2 = [f]+[f^\dag]$, where the annihilation operator of the fermionic mode $f = (c_i+ic_j)/2$. Therefore, errors $[c_i]$ correspond to the exchange of fermions between the system and the environment.

In general, a measurement transforms the input state $\rho_\text{in}$ into the output state
\begin{eqnarray}
\rho_q = \frac{[M_q]\rho_\text{in}}{\Tr ([M_q]\rho_\text{in})},
\end{eqnarray}
where $q$ denotes the measurement outcome and $\sum_q M_q^\dag M_q = \openone$. The measurement outcome is read from the state of the measurement instrument. For a noisy measurement, the system is also coupled to the environment. Because the state of the measurement instrument and the environment may not be fully read, we write the output state of a noisy PP corresponding to the outcome $\mu = \pm 1$ as
\begin{eqnarray}
\rho^{(\mu)} = \frac{\sum_q [M^{(\mu)}_q]\rho_\text{in}}{\Tr (\sum_q [M^{(\mu)}_q]\rho_\text{in})},
\end{eqnarray}
where $\sum_q M^{(+1)\dag}_q M^{(+1)}_q + \sum_q M^{(-1)\dag}_q M^{(-1)}_q = \openone$. We would like to remark that we have assumed that errors are Markovian.

Each operator $M^{(\mu)}_q$ can be expressed as
\begin{eqnarray}
M^{(\mu)}_q = \sum_{\{u_i = 0,1\}}\alpha^{(\mu)}_q (\{u_i\}) c_0^{u_0}c_1^{u_1}c_2^{u_2}c_3^{u_3}.
\end{eqnarray}
We will show that, in $[M^{(\mu)}_q]\rho_\text{in}$, only terms
$$c_0^{u_0}c_1^{u_1}c_2^{u_2}c_3^{u_3}\rho_\text{in}c_0^{v_0}c_1^{v_1}c_2^{v_2}c_3^{v_3}$$
satisfying
\begin{eqnarray}
c_0^{u_0}c_1^{u_1}c_2^{u_2}c_3^{u_3}c_0^{v_0}c_1^{v_1}c_2^{v_2}c_3^{v_3} = \pm 1
\label{eq:condition1}
\end{eqnarray}
or
\begin{eqnarray}
c_0^{u_0}c_1^{u_1}c_2^{u_2}c_3^{u_3}c_0^{v_0}c_1^{v_1}c_2^{v_2}c_3^{v_3} = \pm c_0c_1c_2c_3
\label{eq:condition2}
\end{eqnarray}
can survive, and all other terms vanish after some redundant exchange gates. First, we consider PP on a qubit, i.e.~$c_0,c_1,c_3,c_4$ are respectively $s,x,y,z$ of the qubit. We introduce three operators $Q$, $Q_\text{X}$ and $Q_\text{Z}$, which are TCs of the qubit, a neighbouring red octagon and a neighbouring green octagon, respectively. Because all of these charges are known (the ideal state without any error is a common eigenstate of these three operators), an operation $[A]$ does not change the \textit{state} of the system, where $A = Q, Q_\text{X}, Q_\text{Z}$. Here, the operation $[A]$ can be performed with exchange gates. We would like to remark that, by \textit{state}, we mean the ideal state rather than the physical state of the system, i.e.~the fidelity is not changed by the operation. To remove unwanted terms, each operation $[A]$ is performed with the chance $1/2$ before and after PP. Each random operation can be expressed as $\mathcal{R}_A = ([\openone]+[A])/2$. With these random operations, $[M^{(\mu)}_q]\rho_\text{in}$ is transformed to
$$\mathcal{R}_{Q} \mathcal{R}_{Q_\text{X}} \mathcal{R}_{Q_\text{Z}} [M^{(\mu)}_q] \mathcal{R}_{Q_\text{Z}} \mathcal{R}_{Q_\text{X}} \mathcal{R}_{Q} \rho_\text{in}.$$
As a result, a random phase is added to each term except terms satisfying Eq.~(\ref{eq:condition1}) or Eq.~(\ref{eq:condition2}), and all terms with a random phase are removed. A PP for measuring TC of an octagon is similar. The corresponding three charges are shown in Fig.~\ref{fig:stabilisers}.

After removing unwanted cross terms, the output state can be rewritten as
\begin{eqnarray}
\rho^{(\mu)} \propto \sum_{\nu = \pm 1} \sum_{\{u_i\}}
\epsilon^{(\mu)}_{\nu}(\{u_i\}) [c_0^{u_0}c_1^{u_1}c_2^{u_2}c_3^{u_3}] [\pi_\nu] \rho_\text{in}.
\end{eqnarray}
After symmetrization operations, we can obtain the error model given in the main text. There are two kinds of symmetrization operations. By exchanging four measured MFs, one can obtain the symmetry between these four MFs. By moving a measured MF around an ancillary MF (exchanging them twice), one can change TC of four measured MFs to obtain the symmetry between two outcomes.

\begin{table*}[tbp]
\begin{center}
\begin{tabular}{|c|c|c|c|c|c|c|c|c|}
\hline
$n_\text{D}$ & $p_\text{F}$ & $p_\text{B}$ & $\kappa$ & $\sigma_\kappa$ & $\nu$ & $\sigma_\nu$ & $\eta$ & $\sigma_\eta$ \\ \hline \hline
$0$ & $2\times 10^{-4}$ & $2\times 10^{-4}$ & $1.3728$ & $0.0049$ & $0.6235$ & $0.0186$ & $3.5590$ & $0.0117$ \\ \hline
$0$ & $1\times 10^{-4}$ & $1\times 10^{-3}$ & $1.4594$ & $0.0075$ & $0.8656$ & $0.0264$ & $3.0506$ & $0.0174$ \\ \hline
$0$ & $1\times 10^{-5}$ & $1\times 10^{-3}$ & $1.9831$ & $0.0140$ & $0.7166$ & $0.0437$ & $2.3192$ & $0.0286$ \\ \hline
$0$ & $1\times 10^{-6}$ & $1\times 10^{-3}$ & $2.2076$ & $0.0125$ & $-0.0181$ & $0.0336$ & $2.3652$ & $0.0214$ \\ \hline
$0$ & $1\times 10^{-7}$ & $1\times 10^{-3}$ & $2.2108$ & $0.0252$ & $-0.1064$ & $0.0857$ & $2.3540$ & $0.0560$ \\ \hline
$1$ & $1\times 10^{-4}$ & $1\times 10^{-3}$ & $1.0992$ & $0.0036$ & $0.5564$ & $0.0135$ & $3.5959$ & $0.0070$ \\ \hline
$1$ & $1\times 10^{-5}$ & $1\times 10^{-3}$ & $2.5961$ & $0.0101$ & $0.6387$ & $0.0243$ & $3.5312$ & $0.0161$ \\ \hline
$1$ & $1\times 10^{-6}$ & $1\times 10^{-3}$ & $4.2007$ & $0.0114$ & $0.5441$ & $0.0221$ & $3.5846$ & $0.0113$ \\ \hline
$1$ & $1\times 10^{-7}$ & $1\times 10^{-3}$ & $5.7353$ & $0.0127$ & $0.6551$ & $0.0144$ & $3.3585$ & $0.0078$ \\ \hline
\end{tabular}
\end{center}
\caption{
Parameters $\kappa$, $\nu$ and $\eta$ and their standard deviations.
}
\label{table}
\end{table*}

\begin{figure}[tbp]
\centering
\includegraphics[width=1\linewidth]{\figpath 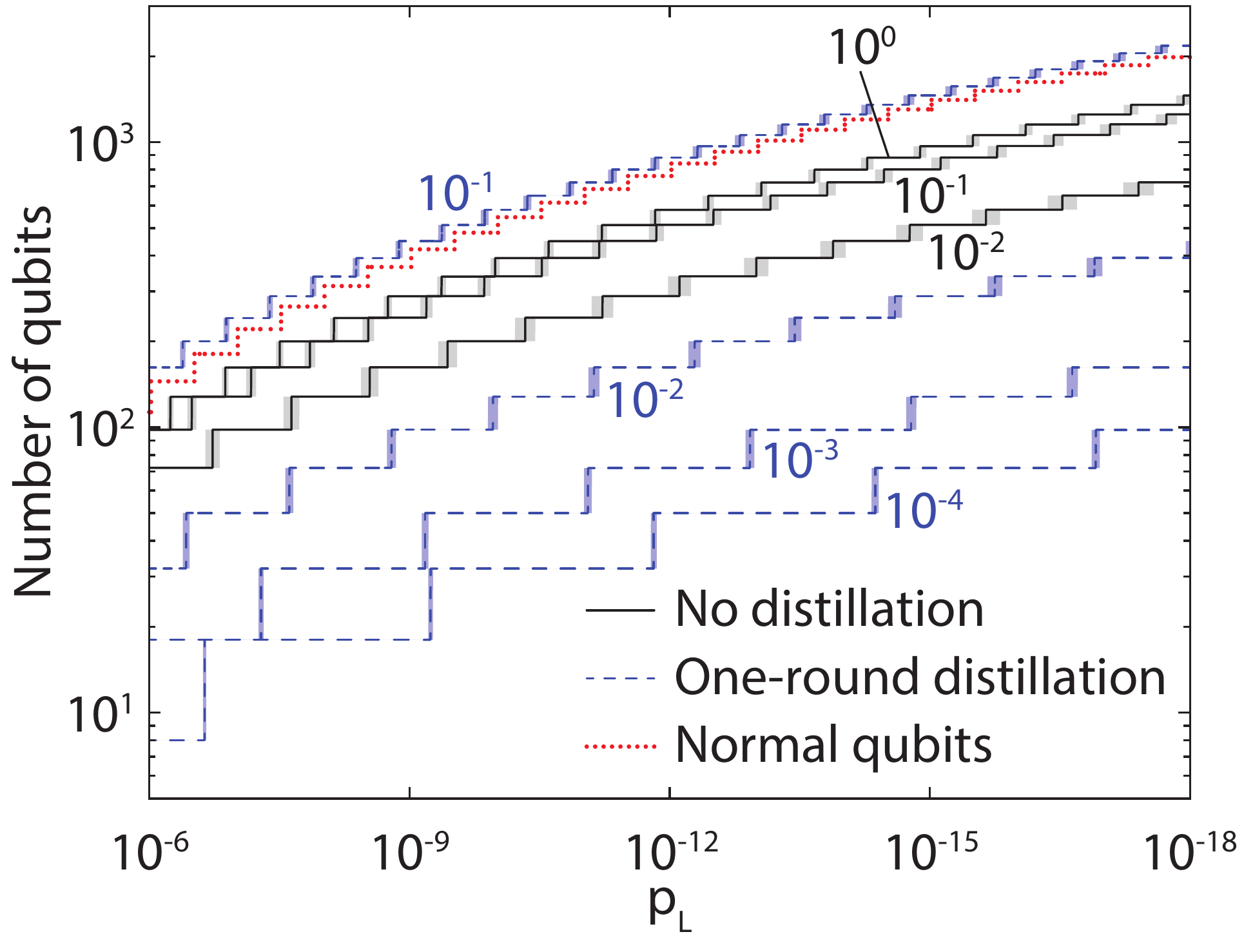}
\caption{
The minimum number of qubits for encoding a logical qubit to achieve a logical-qubit error rate not higher than $p_\text{L}$. The fidelity of PPs is $\sim 99.9\%$, i.e.~$p_\text{B} = 0.02\%$ when $p_\text{F}/p_\text{B} = 1$, and $p_\text{B} = 0.1\%$ when $p_\text{F}/p_\text{B} = 10^{-1},10^{-2},10^{-3},10^{-4}$. For normal qubits, stabiliser measurements are implemented using CNOT gates with the fidelity $99.9\%$~\cite{Fowler2012}. From top to bottom, solid lines respectively correspond to $p_\text{F}/p_\text{B} = 1,10^{-1},10^{-2}$ (further reducing the ratio only slightly reduce the qubit cost), and dashed lines respectively correspond to $p_\text{F}/p_\text{B} = 10^{-1},10^{-2},10^{-3},10^{-4}$. The top line corresponds to $p_\text{F}/p_\text{B} = 10^{-1}$, in which case the distillation is even harmful because the ratio is not low enough. To estimate the number of qubits, we have used Eq.~(\ref{eq:scaling}), where parameters $\kappa$, $\nu$ and $\eta$ are obtained numerically. These parameters and their standard deviations are given in Table~\ref{table}. The uncertainty of the number of qubits due to deviations of parameters is marked with gray and light blue bars: the upper (lower) bound is obtained by replacing $\kappa$, $\nu$ and $\eta$ with $\kappa-\sigma_\kappa$, $\nu-\sigma_\nu$ and $\eta-\sigma_\eta$ ($\kappa+\sigma_\kappa$, $\nu+\sigma_\nu$ and $\eta+\sigma_\eta$).
}
\label{fig:resource_cost}
\end{figure}

\section{The error model and charge tunnelling errors}
\label{app:tunnelling}

There are two kinds of measurement errors: charge tunnelling events (the charge is changed by noise in the measurement operation) and incorrect records (the charge is not changed but the record of the charge is incorrect). When charge tunnelling events are rare, one can repeat the measurement to improve the measurement fidelity. Roughly speaking, by repeating the measurement, measurement errors caused by charge tunnelling events accumulate linearly, but measurement errors caused by incorrect records are suppressed exponentially. Using the measurement with the improved fidelity, a pair of MFs can be initialised (created) with a high fidelity (assuming the measurement is non-destructive). Therefore, we assume that rates of creation and measurement errors are determined by the rate of charge tunnelling events.

In noisy PPs, errors corresponding to $\epsilon_{+,1}$ and $\epsilon_{-,1}$ are charge tunnelling errors, and errors corresponding to $\epsilon_{-,0}$ and $\epsilon_{-,2}$ are measurement errors that do not change TC of four MFs. When charge tunnelling events are rare, TC of four MFs is rarely changed by the noisy PP. Similar to the measurement of two MFs, one can repeat PP to reduce $\epsilon_{-,0}$ and $\epsilon_{-,2}$ errors. Therefore, we assume that $\epsilon_{-,0}$ and $\epsilon_{-,2}$ are determined by the rate of charge tunnelling events. We would like to remark that $\epsilon_{+,2}$ errors do not change TC of four MFs but cannot be detected by repeating PP, and $\epsilon_{+,2}$ errors accumulate approximately linearly when PP is repeated.

When $\epsilon_\text{c} = \epsilon_\text{m} = \epsilon_{+,1} = \epsilon_{-,0} = \epsilon_{-,1} = \epsilon_{-,2} = p_\text{F}$, the total error rate of PPs is at least four times higher than error rates of creation and measurement operations. Such an assumption is reasonable, because usually it is harder to control a quantum system with more subsystems, i.e.~a quantum operation performed on the system with more subsystems usually causes more errors.

\begin{figure*}[tbp]
\centering
\includegraphics[width=0.85\linewidth]{\figpath 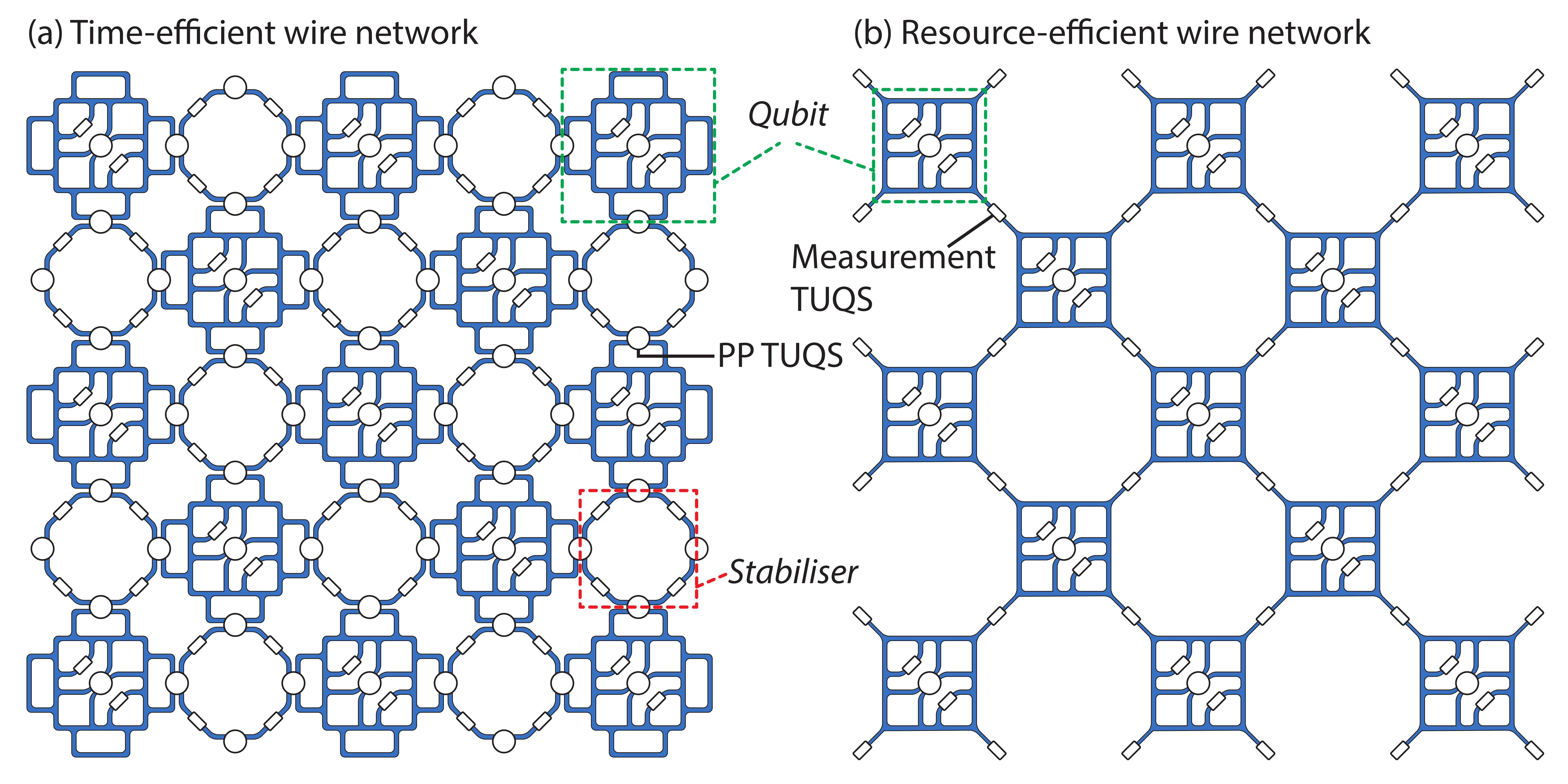}
\caption{
Wire networks for implementing the surface code of MF qubits. Empty rectangles and circles on the wire network respectively denote TUQSs used for measuring TC of two and four MFs (measurements and PPs). In addition to these TUQSs, we also need TUQSs for preparing magic states. Usually, one magic-state TUQS per logical qubit is enough for efficiently preparing logical qubits in magic states~\cite{Li2015}.
}
\label{fig:wire_network}
\end{figure*}

\section{Correction of qubit-charge errors}
\label{app:scheme}

Without correcting qubit-charge errors, Pauli errors are directly corrected. If qubit-charge error are corrected, the error correction of qubit-charge errors is performed before correcting Pauli errors, and then Pauli errors are corrected. When distillations are used, all PPs in stabiliser measurements are replaced by distilled PPs.

The error correction of qubit-charge errors is performed using the lattice shown in Fig.~\ref{fig:lattice}. In (a), each red circle denotes a X stabiliser, each blue circle denotes a qubit. Similarly, in (b), each green circle denotes a Z stabiliser. There are two kinds of charge errors on such lattices. For each qubit, there may be a $[y]$ error, which changes the charge of the qubit. For each stabiliser, there may be an $[a]$ error, which changes the charge of ancillary MFs of the stabiliser. There are four pairs of ancillary MFs for each stabiliser, and we set that the $[a]$ error changes the charge of the pair shared by PPs on qubit-0 and qubit-3 (see Fig.~\ref{fig:lattice}). Errors changing the charge of other pairs are equivalent to combinations of the $[a]$ error and Pauli errors.

When $p_\text{F} = 0$, an $[a]$ error and a $[y]$ error always occur at the same time on a stabiliser and a neighbouring qubit. In Fig.~\ref{fig:lattice}, each edge connecting a stabiliser and a qubit denotes such a correlated error $[ay]$. If $[a]$ corresponds to an X (Z) stabiliser, there is always the Pauli error $[\tilde{Z}_i][\bar{X}_i]$ ($[\tilde{X}_i][\bar{Z}_i]$) occurring simultaneously with the $[ay_i]$ error, where $i=0,1,2,3$ are labels of neighbouring qubits (see Fig.~\ref{fig:lattice}). Here, $[\tilde{A}_i] = [A_0], [A_1], [\openone], [\openone]$ and $[\bar{A}_i] = [\openone], [A_0], [A_3], [\openone]$ when $i=0,1,2,3$. These Pauli errors are corrected once an $[ay]$ error is detected.

When $p_\text{F}$ is finite, $[a]$ errors and $[y]$ errors may occur independently. In Fig.~\ref{fig:lattice}, all empty circles are images of the same vertex corresponding to the \textit{boundary} stabiliser, which is the total TC of all qubits and ancillary MFs. Each edge connecting an X or Z stabiliser and the boundary stabiliser denotes an independent $[a]$ error, and each edge connecting a qubit and the boundary stabiliser denotes an independent $[y]$ error.

Each lattice shown in Fig.~\ref{fig:lattice} corresponds to one round of X or Z stabiliser measurements. When $p_\text{F} = 0$, charge errors occurring in each round of X or Z stabiliser measurements could be corrected independently using the two-dimensional lattice. However, when $p_\text{F}$ is finite, the outcome of the qubit-charge measurement may be incorrect. In this case, charge errors need to be corrected on a three dimensional lattice, on which lattices in (a) and (b) occur alternatively along the third dimension. Between two neighbouring layers, qubits are connected by edges denoting qubit-charge measurement errors (PP for measuring TC of a qubit reports a wrong outcome). Associated with a measurement error, outcomes of corresponding stabiliser measurements are incorrect, which should be corrected once a qubit-charge measurement error is detected.

The weight of an edge on the error correction lattice is given by
\begin{eqnarray}
w = \ln \frac{1-p}{p},
\end{eqnarray}
where $p$ is the rate of errors corresponding to the edge. Correction operations are determined by pairing error syndromes using the Edmonds's minimum weight matching algorithm~\cite{Dennis2002, Kolmogorov2009}.

\begin{figure*}[tbp]
\centering
\includegraphics[width=0.65\linewidth]{\figpath 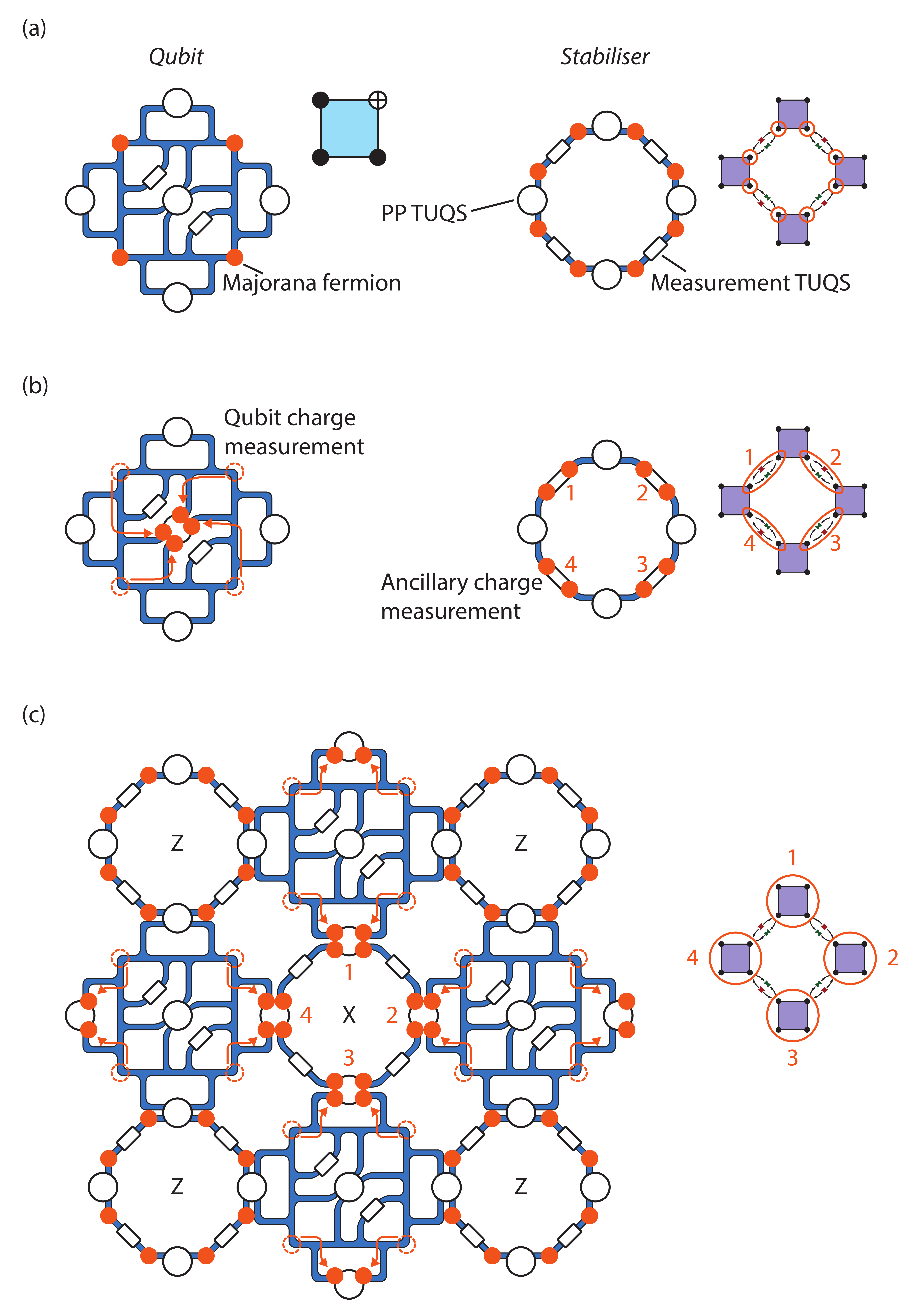}
\caption{
Implementing stabiliser measurements on the wire network. (a) Four MFs are created on each {\it Qubit}, which are used to encode a qubit in the surface code. Eight MFs are created on each {\it Stabiliser}, which are used as ancillary MFs in eight-MF TC measurements. (b) Four MFs on the {\it Qubit} are transferred to the PP TUQS to perform the PP operation. TC of each pair of ancillary MFs are measured using measurement TUQSs on the {\it Stabiliser}, which corresponds to the creation or measurement operation in the eight-MF TC measurement. (c) MFs are transferred to corresponding PP TUQSs connecting {\it Qubits} and {\it Stabilisers} to perform PPs in the eight-MF TC measurement. In the figure, we show the case of measuring X stabilisers. When we perform eight-MF TC measurements corresponding to Z stabilisers, qubit MFs are transferred to the other set of PP TUQSs that connect {\it Qubits} and Z {\it Stabilisers}.
}
\label{fig:implementation}
\end{figure*}

\section{Threshold and resource cost numerical simulations}
\label{app:data}

To estimate thresholds, we calculate rates of logical errors (per round of stabiliser measurements) using the Monte Carlo method. The results are shown in Fig.~\ref{fig:data}. Pauli errors on MF qubits are corrected with a simple cubic lattice and using the Edmonds's minimum weight matching algorithm. Therefore, our error correction protocol is not optimised to correct correlated errors, and one can improve the performance of the code, e.g.~the threshold, by taking correlations into account as proposed in Ref.~\cite{Wang2011}. In our numerical simulations, we has used periodic boundary conditions for simplification, which is sufficient for estimating the threshold. Compared with open boundary conditions, periodic boundary conditions usually result in a lower logical error rate~\cite{Fowler2013b}. If topological qubits allow fault-tolerant quantum state transfer, the code with periodic boundary conditions is feasible in a topological quantum computer.

To estimate the resource cost, another set of Monte Carlo simulations are performed to calculate rates of logical errors shown in Fig.~\ref{fig:scaling}. Because it is hard to directly calculate an extremely low logical error rate using the Monte Carlo method, we use an extrapolation procedure to explore the performance of the code at low logical error rates. The performance of the error correction is determined by the rates of errors in operations used to complete stabiliser measurements, which are the rate of creation errors $p_\text{c}$, the rate of measurement errors $p_\text{m}$, and rates of errors in (distilled) PPs. Therefore, we can use a set of error rates $\bar{p} = (p_\text{c},p_\text{m},p'_{+,1},p'_{+,2},p'_{-,0},p'_{-,1},p'_{-,2})$ to describe errors occurring in stabiliser measurements. Here, $p_\text{c} = p_\text{m} = p_\text{F}$, and $(p'_{+,1},p'_{+,2},p'_{-,0},p'_{-,1},p'_{-,2})$ describe errors in (distilled) PPs. Without the PP distillation, $p'_{+,1} = p'_{-,0} = p'_{-,1} = p'_{-,2} = p_\text{F}$ and $p'_{+,2} = p_\text{B}$; when the distillation is performed, $(p'_{+,1},p'_{+,2},p'_{-,0},p'_{-,1},p'_{-,2})$ can be obtained numerically by simulating the distillation circuit. If error rates are much lower than the threshold or the size of the surface code lattice $d$ is too large, the logical error rate will be extremely low, in which case using the Monte Carlo method to directly calculate the logical error rate is inefficient. Therefore, using the Monte Carlo method, we only calculate logical error rates corresponding to amplified error rates $r\bar{p}$ and small size surface code lattices (see Fig.~\ref{fig:scaling}); and then we fit logical error rates using the function
\begin{eqnarray}
p_\text{L} = \exp[(\alpha_0 + \beta_0 \ln r)d + \alpha_1 \ln d + \alpha_2].
\label{eq:fitting}
\end{eqnarray}
Similar scaling behaviours have been reported in Refs.~\cite{Raussendorf2007NJP, Fowler2013a, Watson2014}. In our case, one can find that this function provides a good fit to our numerical data. Given fitting parameters $\alpha_0$, $\alpha_1$, $\beta_0$ and $\beta_1$, we can estimate the logical error rate for error rates $\bar{p}$ by replacing $r$ with $1$, which gives
\begin{eqnarray}
p_\text{L} = \exp(- \kappa d - \nu \ln d - \eta),
\label{eq:scaling}
\end{eqnarray}
where $\kappa = -\alpha_0$, $\nu = -\alpha_1$ and $\eta = -\alpha_2$. Parameters $\kappa$, $\nu$, $\eta$ and their standard deviations are shown in Table~\ref{table}.

For normal qubits, the rate of errors on a logical qubit in one round of stabiliser measurements is $p_\text{L} \simeq 0.03(p/p_\text{th})^{(d+1)/2}$~\cite{Fowler2012}, where $p$ is the error rate of operations, and $p_\text{th}$ is the threshold. Because $p_\text{th} \simeq 1\%$ and we have chosen $p = 0.1\%$, the cost of normal qubits is estimated using $p_\text{L} = 0.03\times 0.1^{(d+1)/2}$.

Resource costs of different error rates are shown in Fig.~\ref{fig:resource_cost}, which is the full version of Fig.~\ref{fig:plots}(b).

\section{Wire networks}
\label{app:network}

In this section, we discuss how to use the system in Fig.~\ref{fig:wire_code}(a) to implement the surface code. The system corresponding to a $d=3$ surface code is shown in Fig.~\ref{fig:wire_network}(a). The fault-tolerant quantum computing needs a much larger array in order to encode enough logical qubits and perform operations on logical qubits. We also need TUQSs for preparing magic states, which are not shown in Fig.~\ref{fig:wire_network}(a).

To implement the surface code, we need to create four MFs on each {\it Qubit} and eight MFs on each {\it Stabiliser} as shown in Fig.~\ref{fig:implementation}(a). MFs on {\it Qubits} are used to encode qubits, and MFs on {\it Stabilisers} are used as ancillary MFs for performing eight-MF TC measurements [unnumbered MFs in Fig.~\ref{fig:circuit}(a) or $a$ and $b$ MFs in Fig.~\ref{fig:circuit_t}(a)]. Stabiliser measurements are performed in four steps (also see Sec.~\ref{sec:StaMea}):
\begin{itemize}
\item[(i)] Four MFs on each qubit are transferred along the wire network to the PP TUQS of the qubit, and then the TC of four MFs are measured using the PP TUQS [Fig.~\ref{fig:implementation}(b)]. At the same time, TC of each pair of MFs on {\it Stabilisers} is measured using a measurement TUQS on the {\it Stabiliser}, which corresponds to the creation of ancillary MFs with recorded initial TC [creations in Fig.~\ref{fig:circuit}(a) and Fig.~\ref{fig:circuit_t}(a)]. Here, we assume that measurements are non-destructive.
\item[(ii)] MFs are transferred to corresponding PP TUQSs connecting {\it Qubits} and X {\it Stabilisers} to perform PPs [Fig.~\ref{fig:implementation}(c)]. These PPs are a part of the eight-MF TC measurement [four PPs in Fig.~\ref{fig:circuit}(a) and Fig.~\ref{fig:circuit_t}(a)].
\item[(iii)] Operations in step-i are repeated. TC of each qubit is measured again, and TCs of ancillary MFs are measured to complete the eight-MF TC measurement [measurements in Fig.~\ref{fig:circuit}(a) and Fig.~\ref{fig:circuit_t}(a)].
\item[(iv)] To complete one round of stabiliser measurements, a different set of eight-MF TC measurements are performed. The operations are the same as operations in step-ii, but this time Z {\it Stabilisers} instead of X {\it Stabilisers} are involved.
\end{itemize}
In such a system, the state of a qubit can be initialised and measured using measurement TUQSs on the corresponding {\it Qubit}, and one can perform exchange gates using T structures of the wire network~\cite{Alicea2011}. As we have shown, each full round of stabiliser measurements can be performed in 4 steps.

In the system shown in Fig.~\ref{fig:wire_network}(a), there are five PP TUQSs per qubit. The number of PP TUQSs can be reduced using the system shown in Fig.~\ref{fig:wire_network}(b), in which there is only one PP TUQS per qubit. Stabiliser measurements can also be implemented in such a system: all MFs are created on {\it Qubits} (four qubit MFs and eight ancillary MFs per qubit), all PPs are performed using the only PP TUQS of each qubit, and TCs of ancillary MFs are measured using the measurement TUQS connecting two {\it Qubits}. In this resource-efficient system, each full round of stabiliser measurements can be performed in 6 steps. By creating more MFs on each {\it Qubit}, one can also implement the PP distillation in such a system without changing the topology of the network.

\section{Time cost}
\label{app:time}

The time cost depends on the density of TUQSs. We assume one PP TUQS is allocated to each qubit. One round of stabiliser measurements (without distillation) can be completed in $6$ steps. Each distilled PP needs four raw PPs. Therefore, in the case of only one round of distillations, the minimum time cost is $24$ steps per round of stabiliser measurements. Each distilled PP fails with the rate $\sim 4p_\text{B} \simeq 0.4\%$ (assuming $p_\text{F} \ll p_\text{B}$ and $p_\text{B} \simeq 0.1\%$). In order to overcome these failures, usually more qubit resources are needed~\cite{Barrett2010, Whiteside2014}. Alternatively, one can choose to repeat the distillation circuit (the entangled-state generation part, see Fig.~\ref{fig:circuit_t}) for once if a failure occurs. As a result, the failure rate is effectively reduced to $\sim (4p_\text{B})^2 \simeq 1.6\times 10^{-5}$, in which case the qubit cost due to failures should be negligible, but the time cost is doubled.


\begin{thebibliography}{9}
\bibitem{footnote1} For a pair of MFs, $Q$ is the parity of the filling number of the corresponding fermionic mode, i.e.~$Q = -ic_ic_j = 1 - 2f^\dag f$; for $2n$ MFs, similarly $Q$ is the parity of the filling number of the corresponding $n$ fermionic modes.
\bibitem{footnote2} Magic-state preparation errors do not affect the performance of the surface code. Once other errors are corrected, high-fidelity magic states can be prepared using purification protocols. Usually, the fault-tolerance threshold of magic-state errors is higher than the threshold of other errors.
\bibitem{footnote3} A charge-conserving error in the form $[c_i a]$ ($i=0,1,2,3$, and $a$ is an ancillary MF) exchanges TC between MFs $\{c_0,c_1,c_2,c_3\}$ and ancillary MFs. Such an error changes TCs of both $\{c_0,c_1,c_2,c_3\}$ and ancillary MFs.
\end{thebibliography}
\end{document}